\documentclass[default,iicol]{sn-jnl}%
\jyear{2023}%

\theoremstyle{thmstyleone}%

\theoremstyle{thmstyletwo}%

\theoremstyle{thmstylethree}%

\usepackage{siunitx}
\usepackage{mhchem}
\DeclareSIUnit\dBm{dBm}
\DeclareSIUnit\cps{cps}
\DeclareSIUnit\gauss{G}
\DeclareSIUnit\Molar{\textsc{m}}
\usepackage[normalem]{ulem}

\begin{document}

\title[Simultaneous nanorheometry and nanothermometry using intracellular diamond quantum sensors]{Simultaneous nanorheometry and nanothermometry using intracellular diamond quantum sensors 
}

\author[1]{\fnm{Qiushi} \sur{Gu}}
\equalcont{These authors contributed equally to this work.}

\author[1]{\fnm{Louise} \sur{Shanahan}}
\equalcont{These authors contributed equally to this work.}

\author[1]{\fnm{Jack W.} \sur{ Hart}}
\equalcont{These authors contributed equally to this work.}

\author[1]{\fnm{Sophia} \sur{Belser}}

\author[1]{\fnm{Noah} \sur{ Shofer}}

\author*[1]{\fnm{Mete} \sur{ Atature}}\email{ma424@cam.ac.uk}

\author*[1]{\fnm{Helena S.} \sur{ Knowles}}\email{hsk35@cam.ac.uk}

\affil[1]{\orgdiv{Cavendish Laboratory}, \orgname{University of Cambridge}, \orgaddress{\street{JJ Thompson Avenue}, \city{Cambridge}, \postcode{CB3 0HE}, \country{United Kingdom}}}


\abstract{Viscoelasticity of the cytoplasm plays a critical role in cell morphology and division. In parallel, local temperature is coupled to viscoelasticity and influences cellular bioenergetics. Probing the interdependence of intracellular temperature and viscoelasticity provides an exciting opportunity for the study of metabolism and disease progression. Here, we present a dual-mode quantum sensor, capable of performing simultaneous nanoscale thermometry and rheometry in a dynamic cellular environment. Our technique uses nitrogen-vacancy centres in nanodiamond, combining sub-diffraction resolution single-particle tracking in a fluidic environment with optically detected magnetic resonance spectroscopy. We demonstrate nanoscale sensing of temperature-dependent viscoelasticity in complex media. We then use our sensor to investigate the interplay between intracellular forces and cytoplasmic rheology in live cells, revealing details of active trafficking and nanoscale viscoelasticity.
}



\maketitle
\section{Main}\label{sec1}

Nanorheology addresses the question of how soft materials deform and flow at the nanoscale \cite{Squires2010, Waigh2016}. Of significant interest in nanorheology is the study of complex cellular media such as the cytoplasm, which heavily influence cellular processes such as transport \cite{GUO2014822}, division \cite{Hurst2021, adeniba2020simultaneous} and morphological changes \cite{Pittman2022}. These properties, like many others in the cell, are linked to local biochemical energetics where temperature plays a critical role \cite{Postmus2008,Kieling2013}. It is well-established that cells regulate their viscoelastic properties in response to external temperature changes through homeoviscous adaption \cite{Sinensky1974, Budin2018} and viscoadaption \cite{Persson2020}. Variations in intracellular temperature, rheology and their interdependence at the nanoscale remain outstanding questions today \cite{Baffou2014, jawerth2018salt} in the pursuit of a deeper understanding of cellular homeostasis, disease progression \cite{Chung2022} and pathways for cancer treatment \cite{sharma2019nanoparticles}. The current challenges for existing biosensing tools include small length scales and poor signal-to-noise ratio of the phenomena under investigation.

Optical techniques can provide means for investigating intracellular phenomena at the nanoscale in a non-invasive way. These methods are often susceptible to variations in autofluorescence \cite{Arai2015}, spectral transmission \cite{Brites2016} and refractive index \cite{Okabe2012}, which are typically present in complex biochemical environments. The interdependence of physical properties in biological systems can also be obfuscated by local inhomogeneity. Further, a change in one property, for example temperature, can often affect others such as viscosity, the speed of chemical reactions or the rate of cell division. The relationship between two properties is thus hard to capture effectively if the level of an external perturbation cannot be measured accurately and independently. Multimodal sensors offer the opportunity to reveal such interdependence. 

Among the many approaches to nanoscale sensing in biological systems that are currently being explored, nanoparticles provide a platform which enables robust optical intracellular sensing \cite{choi2020probing,Brites2016}. Nanodiamonds containing nitrogen-vacancy centres (NV) are one of the leading candidates: their properties include stable photoluminescence (PL), minimal cytotoxicity at high concentrations \cite{Vaijayanthimala2009,woodhams2018graphitic}, amenability to surface functionalisation \cite{Krueger2012}, and robustness against changes in pH \cite{fujiwara2019monitoring}. The ground-state spin transition which is utilised for sensing can be effectively uncoupled from background fluorescence fluctuations, enabling NV measurements to be unaffected by local changes in the optical environment. The NV has the capability to measure several different quantities, as demonstrated separately for temperature  \cite{Neumann2013}, magnetic field \cite{Horowitz2012}, electric field \cite{Bian2021}, pressure \cite{Ho2020}, reactive oxygen species \cite{Nie2021} and through targeted surface functionalisation, pH \cite{Rendler2017}. These demonstrations position the NV as a promising candidate for multi-modal sensing implementations.

In this work, we perform nanothermometry and nanorheology using optically detected magnetic resonance (ODMR) and particle tracking of NV-containing nanodiamonds. We first demonstrate the operational protocol and achieve 3.7-nm spatial resolution with 9.6-ms update rate and a temperature sensitivity of $2.3\,  \mathrm{^{\circ} C/\sqrt{Hz}}$. We quantify the performance in multiple well-controlled fluidic environments and then employ our sensor inside live human cancer cells and and reveal different regimes of intracellular dynamics, while simultaneously measuring temperature. This dual-modality sensing is performed on a custom biosensing chip, capable of microscopic temperature control and coherent spin manipulation.

\begin{figure*}
\includegraphics[width=1\textwidth]{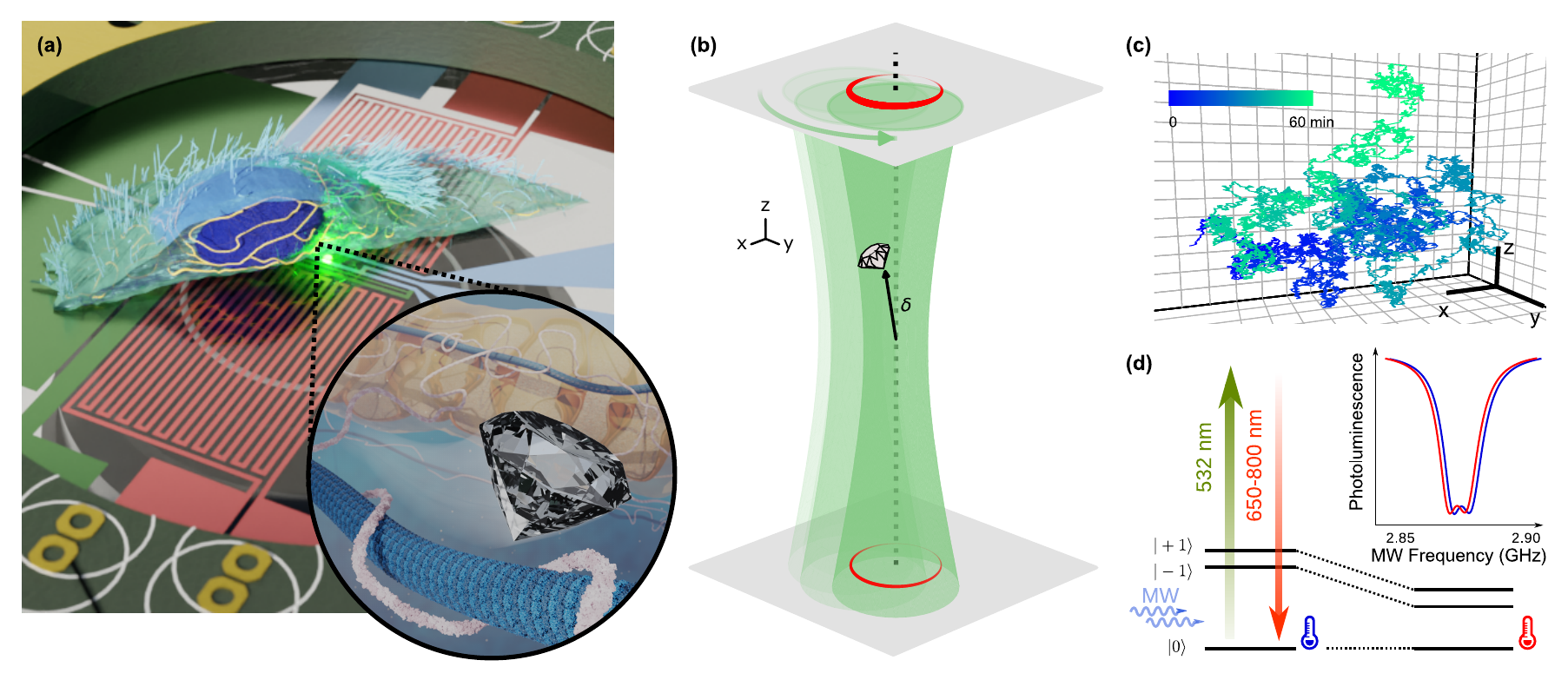}
\caption{\textbf{Diamond-based nanothermometer and nanorheometer.} \textbf{(a)} An illustration of the cross-section of a cell grown on a custom sensing chip, consisting of a resistive temperature detector (blue), two resistive heaters (red) and a coplanar waveguide (green), which is used for accurate temperature control and microwave delivery. \textbf{Inset:} A nanodiamond interacts with its complex surroundings in the cytoplasm, including the microtubules (blue), actin filaments (pink) and mitochondria (yellow in background). \textbf{(b)} Real-time tracking is achieved by collecting the PL from a nanodiamond at two axially offset planes separated by $100\,\mathrm{nm}$ (red) as the excitation laser (green) orbits the last inferred position of the nanodiamond with a radius of $50\,\mathrm{nm}$. Corrections ($\delta$) in the transverse and axial directions are made to counteract any imbalance in PL along the orbit (indicated by the intensity of the red orbit) and between the top and bottom imaging planes (grey shaded planes). \textbf{(c)} An example trajectory of a nanodiamond undergoing Brownian motion in glycerol. The background grid has a spacing of $1\,\mathrm{\mu m}$. \textbf{(d)} The transition frequencies of the NV ground state are temperature dependent and probed using ODMR (top right), with the central frequency of the ODMR spectrum decreasing with increasing temperature (blue to red).
\label{Fig1}}
\end{figure*}

\section{Calibrating the sensor performance of nanodiamonds}
To achieve optical readout of the NV spin, which underlies the sensing concept of nanodiamonds, we use a home-built confocal microscope (Supplementary Information Section 1). Figure ~\ref{Fig1} \textbf{(a)} and its inset illustrate the experimental arrangement, where a nanodiamond moves inside a cell while sensing local temperature. We use nanodiamonds that contain an ensemble of 100-300 NVs, with a radius of ${\sim}$25 nm. Nanoparticles of comparable sizes move with diffusion coefficients exceeding $3\times 10^4\,\mathrm{nm^2\, s^{-1}}$ in cells \cite{Li2018} (Supplementary Information Section 2). These dynamic environments require the nanodiamond to be tracked throughout optical spin readout measurements. We achieve this through a double-plane orbital tracking method \cite{Katayama2009}, which provides real-time feedback control of the nanodiamond’s location, as illustrated in Fig. ~\ref{Fig1} \textbf{(b)}. The excitation laser performs circular orbits in the transverse plane with a period of $9.6\,\mathrm{ms}$. The two confocal collection planes are offset symmetrically by ${\sim}50\,\mathrm{nm}$ in opposite axial directions from the laser focus and collect the NV PL along the two offset circular paths. The asymmetries in PL around the orbit and between the top and bottom planes provide feedback parameters in the transverse and axial directions respectively, updating the centre of the orbital tracking to the nanodiamond position (Supplementary Information Section 3). In Fig. ~\ref{Fig1} \textbf{(c)} we demonstrate the tracking of such a nanodiamond diffusing in glycerol.

Whilst tracking the nanodiamond in real time, we simultaneously perform continuous-wave ODMR for temperature sensing. The ground state zero-field splitting of the NV can be optically read out by driving the NV spin from the $m_s = 0$ state to the $m_s = \pm1$ states. On resonance, this leads to a decrease in PL, as shown in Fig. ~\ref{Fig1} \textbf{(d)}. These transition frequencies are dependent on temperature. We sweep the microwave frequency over the target range every ${\sim}1\, \mathrm{ms}$ and monitor the NV PL continuously to identify the spin resonances. We infer the change in temperature from the change in central frequency of the full ODMR spectrum. The central frequency is extracted using an interpolation method (Methods, Supplementary Information Section 4).

The ODMR-based thermometry technique requires the delivery of microwaves to the region of interest. This typically leads to heating of the substrate and the intracellular medium. These effects can be a challenge to control and may vary from sample to sample. To achieve reproducible temperature control, sample heating with minute-scale temporal resolution (Supplementary Information Section 5) and microwave delivery for the manipulation of NV spins, we developed a custom fabricated chip. 
All measurements are performed using a gold-patterned glass coverslip conprising a coplanar waveguide, two resistive heaters and a resistive temperature detector (RTD), as highlighted in Fig. ~\ref{Fig1}  \textbf{(a)} with green, red and blue regions, respectively. A polydimethylsiloxane (PDMS) open-top well is incorporated into the sensing chip to contain liquid samples when necessary.

\begin{figure*}
\includegraphics[width=1\textwidth]{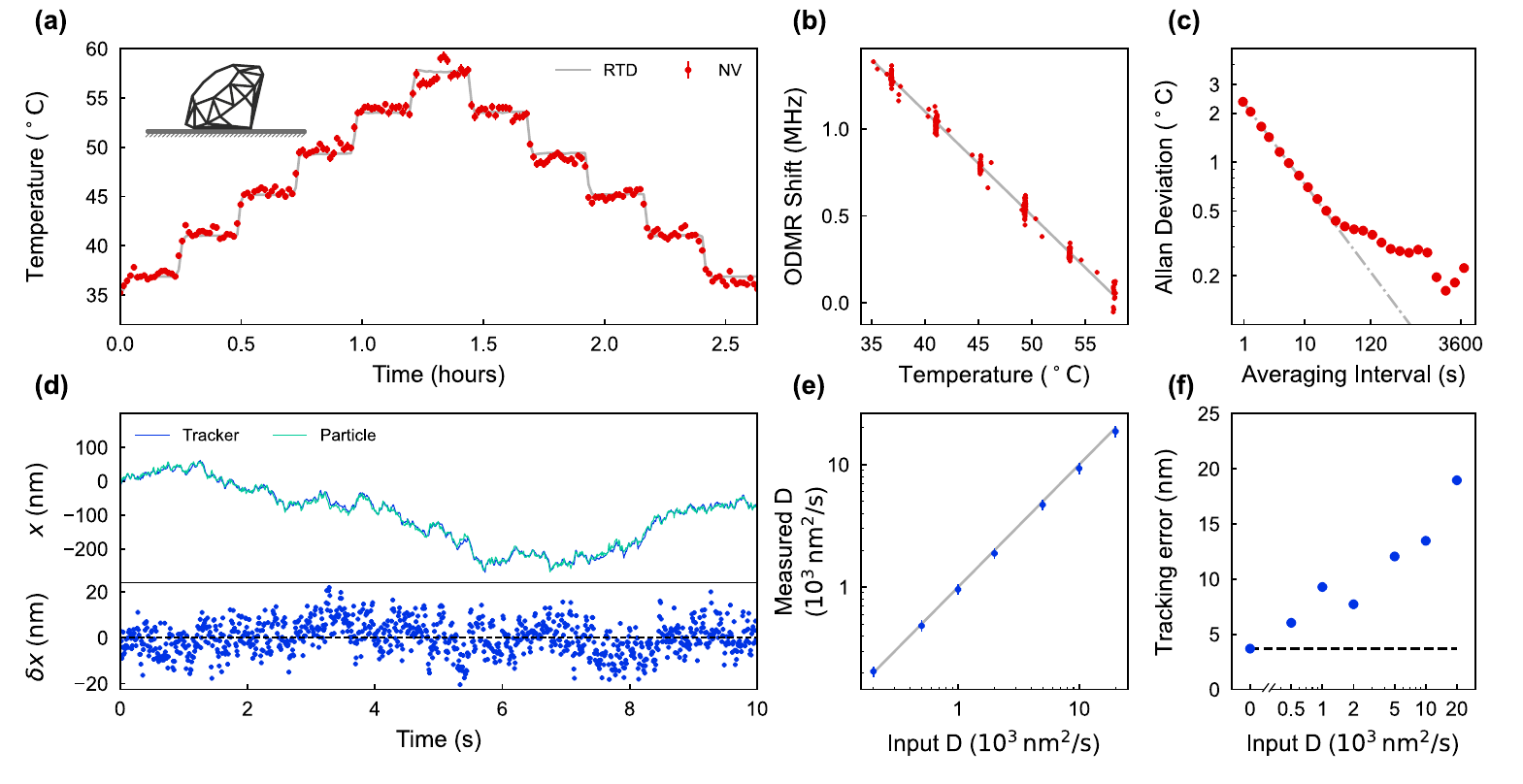}
 \caption{\textbf{Accuracy and precision of the nanodiamond nanothermometer and nanorheometer.} \textbf{(a)} The substrate temperature (grey curve) is stepped by $4\,\mathrm{^{\circ} C}$ every 15 minutes, with the corresponding temperature reported by NV ODMR (red data). \textbf{(b)} The frequency shift is proportional to the change in temperature, with a temperature dependence of $\kappa = -60.0(4)\,  \mathrm{kHz/^{\circ} C}$. \textbf{(c)} The temperature precision over an accumulation time is characterised by the Allan deviation, from which we extract a sensitivity of $2.3\,  \mathrm{^{\circ} C/\sqrt{Hz}}$. \textbf{(d)} Comparison between the known position (cyan data) in the x-direction of a nanodiamond moved in a Brownian motion-manner and the tracker-reported position (blue data), with the corresponding difference ($\delta x$) shown in the lower panel. The set diffusion coefficient is $2\,\times10^3 \mathrm{nm^2/s}$ for this measurement. \textbf{(e)} The measured diffusion coefficient using the mean square displacement (MSD) at a time interval of 1 s shows a close agreement with the input diffusion coefficient. \textbf{(f)} The dynamic tracking accuracy depends on the diffusion coefficient. When the particle is stationary, our system has a benchmark spatial resolution of  3.7-nm with 9.6-ms update rate (black dashed curve). }
\label{Fig2}
\end{figure*}

To benchmark the thermometry modality, we first quantify the temperature sensitivity of a stationary nanodiamond dropcast on the quantum sensing chip in the absence of any fluidic environment. As seen in Fig. ~\ref{Fig2} \textbf{(a)}, the substrate temperature is adjusted in steps of $4\,\mathrm{^{\circ}C}$ and the ODMR central frequency shifts proportionally. We extract a temperature dependence of $\kappa = -60.0(4)\,  \mathrm{kHz/^{\circ} C}$ as shown in Fig. ~\ref{Fig2} \textbf{(b)}. Consistent with previous results, our experiments show that this value can vary between nanodiamonds in the range $ -53.6(1.0)\,  \mathrm{kHz/^{\circ} C} \leq \kappa \leq -91.0(1.0)\,  \mathrm{kHz/^{\circ} C}$ and therefore needs to be calibrated for every nanodiamond (Supplementary Information Section 6).
Using the Allan deviation, Fig. ~\ref{Fig2} \textbf{(c)} shows an extracted sensitivity of $2.3\,\mathrm{^{\circ} C/\sqrt{Hz}}$ which agrees with the shot noise-limited sensitivity as predicted by the Cramer-Rao bound, $2.1\,\mathrm{^{\circ} C/\sqrt{Hz}}$, to within 10 \% (Supplementary Information Section 7).

To benchmark the rheometry modality, we start by verifying the dynamic tracking accuracy of the single particle tracking method. The scanning mirrors and the objective lens are moved such that a stationary nanodiamond on the substrate exhibits a predefined trajectory that mimics Brownian motion (Methods, Supplementary Information Section 8). Figure ~\ref{Fig2} \textbf{(d)} demonstrates the difference between the predefined particle trajectory (blue data) and the tracker readout (cyan data) over a 10 min interval which is used to determine the tracking accuracy. To analyse the stochastic diffusive motion, we compute the 2D mean square displacement (MSD), $\mathrm{MSD}(\tau) = \langle \lvert \mathbf{{r}}(t+\tau)-\mathbf{{r}}(t)\rvert ^2\rangle$, where $\mathbf{r}$ is the position vector in the transverse plane and $\tau$ is the time interval. The MSD depends linearly on the time interval for a particle undergoing Brownian motion, as $\mathrm{MSD} = 4D\tau$, where $D$ is the diffusion coefficient. The measured diffusion coefficient agrees with the input diffusion coefficient as highlighted in Fig. ~\ref{Fig2} \textbf{(e)}. In our system, we reach an upper bound of $D = 5\times10^4\, \mathrm{nm^2\, s^{-1}}$, exceeding the typical intracellular diffusion coefficients observed with similarly sized nanodiamonds (Supplementary Information Section 2). When the particle is stationary, we measure a resolution of $3.7\,\mathrm{nm}$ with a 9.6-ms update rate, as shown in Fig. ~\ref{Fig2} \textbf{(f)}, which is ${\sim}$60 times smaller than the 250-nm radius defined by the $1/e^2$ point-spread function. Our particle tracking is capable of following nanodiamonds in a range of dynamic environments with high enough velocity and spatial resolution to allow the extraction of viscoelastic moduli. The nanodiamonds simultaneously operate as quantum sensors for temperature without the need for measurement deadtime in either modality.

\section{Dual-modality nanosensing in a viscosity-tuneable fluid}

From the stochastic motion of nanoparticles we infer properties about the surrounding material using passive nanorheometry. This provides a quantitative description of the relationship between the nanodiamond motion and the external forces. To demonstrate the use of nanorheometry with simultaneous nanothermometry, we study nanodiamonds undergoing Brownian motion in glycerol. We choose glycerol as it can be assumed homogeneous and predominantly viscous and has a known temperature-dependent viscosity \cite{Volk2018}. In Fig. ~\ref{Fig3} \textbf{(a)} a particle is shown to travel several micrometers in 96 seconds. The particle motion is random, and thus we use the MSD to extract the diffusion coefficient, $D$. 

In glycerol $D$ obeys the Stokes-Einstein relation, $D=\frac{k_\mathrm{B} T}{6\pi r \eta(T)}$, where $T$ is the absolute temperature, $k_\mathrm{B}$ is the Boltzmann constant, $r$ is the hydrodynamic radius of the nanoparticle and $\eta(T)$ is the temperature dependent viscosity. We study the temperature dependence of the diffusion coefficient over a $17.5\,\mathrm{^{\circ} C}$ range by increasing and decreasing the temperature in steps of $3.5\,\mathrm{^{\circ} C}$ every 5 minutes. In the case of glycerol, $\eta(T)$ is linearly dependent on temperature in the range probed, $\eta(T) = \eta _0+\mu (T-T_0 )$, with $\mu =0.0208\, \mathrm{Pa\cdot s/^{\circ} C}$, $T_0=35\mathrm{^{\circ} C}$ and $\eta_0=0.301\, \mathrm{Pa\cdot s}$. \cite{Volk2018}. From the experimental measurements of the diffusion coefficient in Fig. ~\ref{Fig3} \textbf{(b)} (red data), we extract the temperature dependence of the viscosity, $\eta(T)$ (black solid curve), using only the radius of the particle as a fitting constant. The estimated hydrodynamic radius of the nanodiamond is $28(1) \,\mathrm{nm}$, which agrees with the nominal distribution provided by the supplier, $25\,\mathrm{nm}$. As one would intuitively expect, a proportion of the diffusion coefficient's temperature dependence can be attributed to thermal agitation alone as seen in Fig. ~\ref{Fig3} \textbf{(b)} (grey dashed curve).

Figure ~\ref{Fig3} \textbf{(c)} and \textbf{(d)} display the measured temperature, extracted from ODMR, and the diffusion coefficient, extracted from the nanodiamond trajectory, respectively. These nanoscale measurements were verified with the RTD-measured temperature and the corresponding extracted diffusion coefficient respectively (grey curves). Using the multimodal sensor, we are able to probe the link between viscosity and temperature in glycerol through two simultaneous and independent measurements. 

\section{Revealing temperature-dependent viscoelasticity of a complex medium}

\begin{figure*}
\includegraphics[width=1\textwidth]{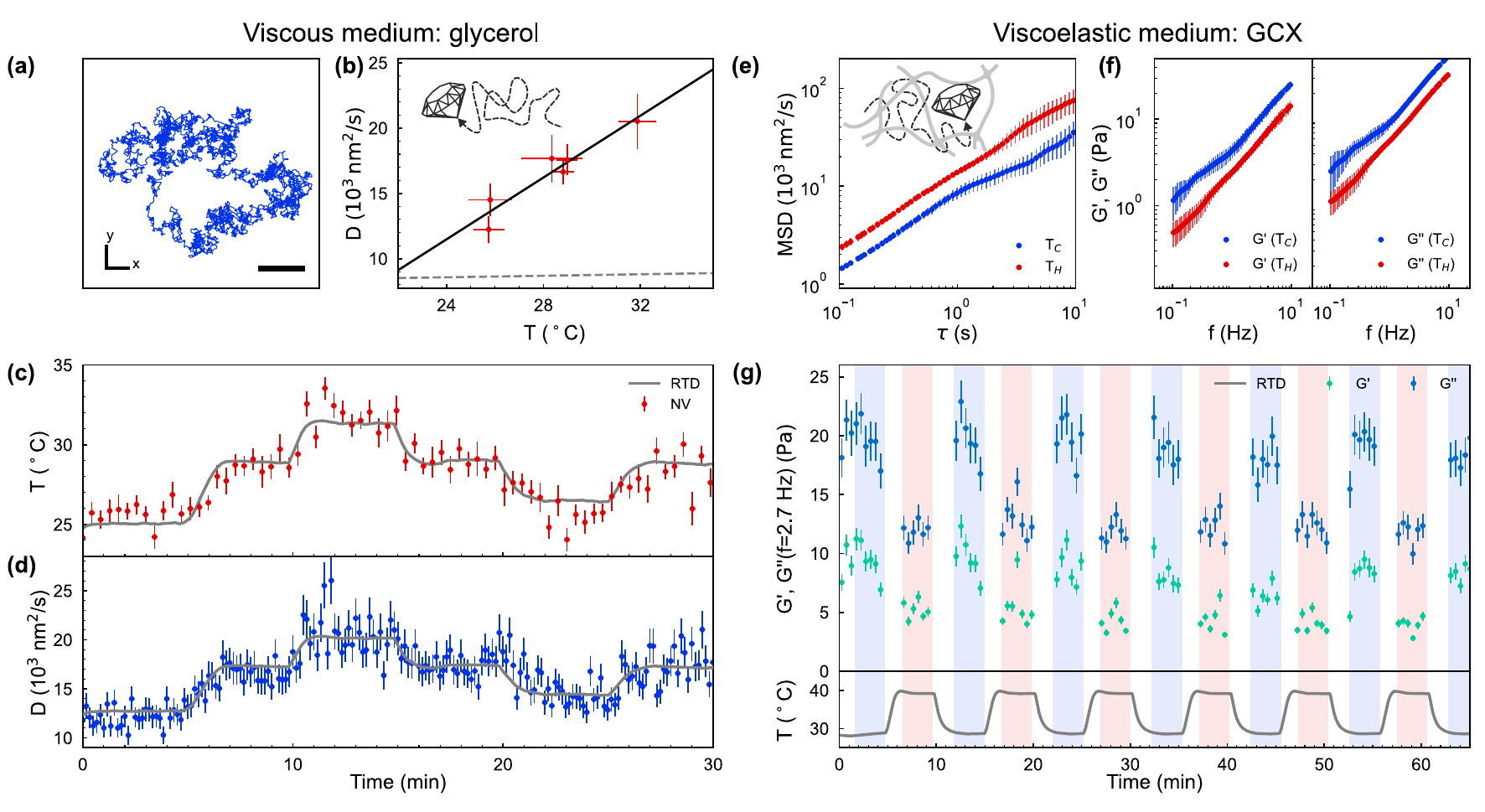}
\caption{\textbf{Temperature and rheology measurements in abiotic media.} \textbf{(a)} An example of the nanodiamond trajectory projected onto the transverse plane over 96 s in glycerol. Scalebar: $1\,\mathrm{\mu m}$. \textbf{(b)} The diffusion coefficient measured at different temperature values (red), with a linear fit (solid black curve) from which a hydrodynamic radius of $28(1)\,\mathrm{nm}$ is extracted. The grey dashed line shows the temperature dependence of the diffusion coefficient assuming a fixed viscosity of $0.919\,\mathrm{Pa\cdot s}$ corresponding to glycerol at $21\,\mathrm{^\circ C}$.\textbf{(c, d)} The simultaneous determination of temperature (red circles) and viscosity (blue circles) in glycerol, a purely viscous medium. The grey curve in (c) shows the temperature read out by the sensing chip and (d) shows the corresponding diffusion coefficient using the radius extracted from (b). \textbf{(e, f)} The mean square displacement (MSD), and viscous ($G^{\prime \prime}$) and elastic ($G^{\prime}$) moduli in a viscoelastic medium, glycerol-crosslinked xanthan (GCX), at $T_\mathrm{C} = 28.7\,\mathrm{^{\circ} C}$ (blue circles) and $T_\mathrm{H}= 39.3\,\mathrm{^{\circ} C}$ (red circles) obtained from nanodiamond tracking. \textbf{(g)} Temperature dependence of $G^{\prime}$  and $G^{\prime \prime}$ at $f =  2.7\,\mathrm{Hz}$ for alternating temperatures $T_\mathrm{C}$ (blue shaded) and $T_\mathrm{H}$ (red shaded) as measured by the sensing chip (grey curve). (e and f) are calculated from the first 3 minutes of data at $T_\mathrm{C}$ and $T_\mathrm{H}$ in (g) (first blue and first red shaded regions).}
\label{Fig3} 
\end{figure*}

In addition to sensing predominantly viscous rheological behaviour, our probe can reveal the viscoelastic properties of biological environments such as DNA hydrogels \cite{Xing2018} and the actin cytoskeleton \cite{Levin2021}. To model these environments, we use the synthetic viscoelastic polymer network glycerol cross-linked xanthan (GCX). The complex modulus, $G^{\ast}(f)=G^{\prime}(f)+iG^{\prime \prime}(f)$, is used to characterise viscoelastic materials and can be calculated from the MSD \cite{PhysRevLett.74.1250,Mason2000}, where $f$ is the frequency of external perturbations at which G is measured. The real part of the complex modulus, $G^{\prime}(f)$, is a measure of the elasticity of the material and the imaginary part, $G^{\prime \prime}(f)$, is a measure of the viscous component. The ratio of the real and imaginary parts of the complex modulus establish whether an environment is dominated by viscosity or elasticity. This capability can be used to capture how the rheological properties of the medium reacts to external perturbations.

Figure ~\ref{Fig3} \textbf{(e)} displays the temperature-dependent MSD, obtained from the nanodiamond single-particle trajectory. In Fig. ~\ref{Fig3} \textbf{(f)} we demonstrate that $\lvert G^{\ast}\rvert$, as well as its real and imaginary components decrease with temperature. This is the expected behaviour for viscoelastic materials from the time-temperature superposition principle \cite{christensen2012theory}, as previously observed in other materials like hydrogels \cite{hao2011viscoelastic}. We achieve sufficient sensitivity to distinguish between the two viscoelastic states with a 30-s averaging interval when we cycle the temperature by $10.6\,\mathrm{^{\circ} C}$. Figure ~\ref{Fig3} \textbf{(g)} presents the change in viscous (blue) and elastic (cyan) moduli at two distinct temperature values of $28.7\,\mathrm{^{\circ} C}$ and $39.3\,\mathrm{^{\circ} C}$.

\section{Capturing signatures of external forces in live cells}

\begin{figure*}
\includegraphics[width=1\textwidth]{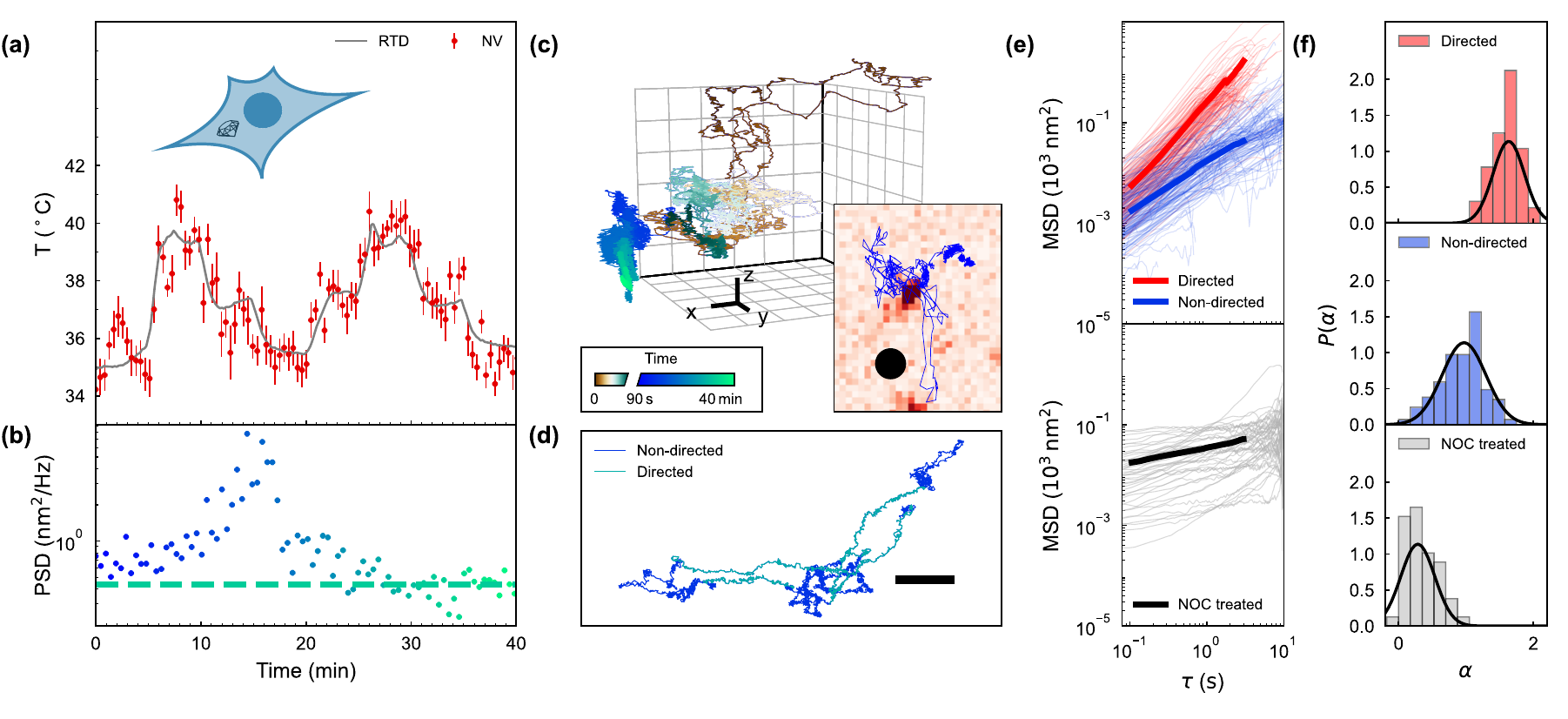}
\caption{\textbf{Nanodiamond multimodal sensing in live cells} \textbf{(a, b)} Simultaneous readout of temperature and power spectral density (PSD) in a cell. The grey curve in (a) shows the temperature read out by the sensing chip and the PSD in (b) corresponds to $f = 40\,\mathrm{Hz}$. The dashed line represents the upper bound of the thermal contribution to the PSD. \textbf{(c)} Trajectory of a nanodiamond in a cell over 40 min. xyz scalebar: $250\,\mathrm{nm}$ \textbf{Inset:} xy particle trajectory relative to the optical diffraction limit (black spot, diameter = $500\,\mathrm{nm}$). \textbf{(d)} xy particle trajectory showing both non-directed (dark blue) and directed (light blue) motion. Scalebar = $500\,\mathrm{nm}$.  \textbf{(e)} The mean square displacement, MSD and ensemble averages (thick lines) for nanodiamonds with non-directed motion (blue) and directed motion (red) in untreated cells, and motion in cells treated with $50\,\mathrm{\mu M}$ nocodazole for 1 hour (grey). \textbf{(f)} Probability densities for the power-law exponents, $\alpha$, for directed motion (red) and non-directed motion (blue) in untreated cells and cells which had been treated with $50\,\mathrm{\mu M}$ nocodazole for 1 hour (grey). The black curves show respectively fitted normal distributions.}
\label{Fig4}
\end{figure*}

Having benchmarked our dual-modal sensing approach in controlled environments, we next investigate the intracellular response to external temperature changes and the motion of nanodimamonds inside cells. We incubate HeLa cells with nanodiamonds and confirm internalisation using 3D confocal microscopy (see Methods and Supplementary Information Section 9). We expose the nanodiamond-containing cells to temperature cycles of $5.0\,\mathrm{^\circ C}$ in steps of $2.5\,\mathrm{^\circ C}$ lasting 5 minutes each. Figure ~\ref{Fig4} \textbf{(a)} confirms the agreement of cell temperature measured independently by nanodiamonds (red circles) and the RTD temperature sensor on the sensing chip (gray curve). 

Unlike glycerol and GCX, the cytoplasm of a cell is an active medium which is neither spatially homogeneous nor in thermal equilibrium. Molecular motors cause collective agitation of the cytoplasm \cite{nishizawa2017feedback, Katrukha2017, GUO2014822, mizuno2007nonequilibrium, mizuno2009high}. As such, particle motion represents the combined effect of both the material properties and the cellular activity. The power spectral density (PSD) of a particle's location, $\langle x^2(\omega)\rangle$, which is the Fourier transform of the MSD, can be used to model this behaviour. For media where the force-displacement relation is linear, this PSD is related to the power spectra of thermal stochastic forces, $\langle \xi^2 (\omega)\rangle \propto k_\mathrm{B} T$, \cite{GUO2014822, nishizawa2017feedback} and external forces due to cell agitation, $\langle F_\mathrm{ext}^2\rangle$, by Hooke's law,

\begin{equation}\label{HookesLaw}
    \vert K(\omega) \vert^2 \langle x^2(\omega)\rangle=\langle \xi^2 (\omega)\rangle + \langle F_\mathrm{ext}^2\rangle .
\end{equation}

\noindent Here, $K(\omega) = (6\pi r ) G^*(\omega) $ is the (complex) spring constant characterising the property of the medium, $G^{\ast}(\omega)$ is the complex modulus introduced in the previous section and $\omega$ is the angular frequency corresponding to the linear frequency, $f$. Figure ~\ref{Fig4} \textbf{(b)} presents the PSD of the nanodiamond location averaged over 30 seconds at $f = 40\,\mathrm{Hz}$. The PSD increases dramatically approximately 15 minutes into the measurement, for a duration of around 10 minutes. Variations in the PSD can be explained by a combination of changes in $\vert K(\omega) \vert$ and $\langle F^2_\mathrm{ext} (\omega)\rangle$. Particular biological events such as cell division can result in large changes in viscoelasticity \cite{Hurst2021, adeniba2020simultaneous} and thus changes in $\vert K(\omega) \vert$. In the absence of such events, active microrheology \cite{GUO2014822,guo2013role} and whole-cell AFM \cite{cai2017temporal} experiments suggest that cell viscoelasticity remains constant over the time scale of hours. The changes we observe are therefore likely dominated by the external forces.

Nanodiamond internalisation involves the endocytic pathway \cite{chu2015rapid} and therefore single-particle trajectories are expected to show active trafficking together with the Brownian motion of the particle. As the nanodiamond spends the majority of time in Brownian motion, the time-averaging used in the MSD and PSD analysis can hide transient features in the trajectory. Figure ~\ref{Fig4} \textbf{(c)} shows the full trajectory of a nanodiamond in a cell over 40 minutes. We analyse this data by categorising segments of nanodiamond trajectories according to periods of statistically significant directional persistence, characterised by the directionality ratio, $\gamma = d/l$, where $d$ and $l$ are the displacement and distance of a trajectory portion respectively (Supplementary Information Section 10). Figure ~\ref{Fig4} \textbf{(d)} shows an example of a nanodiamond trajectory containing directed motion segments. We compare the results from our segmentation method with the spread of anomalous diffusion exponents, $\alpha$. Through the relation $\mathrm{MSD}\propto \tau^\alpha$, the displacement behaviour is typically classified into subdiffusive ($\alpha<1$), diffusive ($\alpha=1$) and superdiffusive ($\alpha>1$) states. Separating the trajectories into segments reveals that when the nanodiamonds are not in the directed motion state, they on average exhibit Brownian-like behaviour, as can be seen from the MSDs in Fig. ~\ref{Fig4} \textbf{(e, top - blue)} resulting in a power-law exponent of 0.97(5) in Fig. ~\ref{Fig4} \textbf{(f, middle)}. In comparison, Fig. ~\ref{Fig4} \textbf{(e, top - red)} and \textbf{(f, top)}, show that the nanodiamonds in the directed motion state appear superdiffusive, with a power-law exponent of 1.65(5). The directed motion of the nanodiamonds could represent active trafficking around the cell interior. To investigate the effect of molecular motors, 50 $\mu$M of nocodazole was added to destabilise the microtubule network \cite{Li2018}. Under this treatment, nanodiamond trajectories exhibited no directed motion. Further, the average power-law exponent of 0.3(1) indicates subdiffusive motion as presented in Fig. ~\ref{Fig4} \textbf{(e, bottom)} and \textbf{(f, bottom)}. From this we can infer that in the absence of external forces caused by microtubule-associated processes, the cytoplasm behaves as an elasticity-dominated weak gel \cite{GUO2014822,deng2006fast} (Supplementary Information Section 12).

\section{Conclusions}
Multimodal quantum sensing opens new avenues for investigating perturbation and response accurately and independently at the nanoscale in active biological environments. We identify directed motion of the nanodiamonds as a possible indicator of active trafficking in the cell. Further, by removing the action of the molecular motors, we show that the cytoplasm is dominated by its elastic properties. Our results also show that within our measurement sensitivity and at the nanodiamond position, HeLa cells do not regulate their internal temperature in the presence of an external thermal perturbation (Supplementary Information Section 11).

The orbital tracking method we employ is not limited to the continuous-wave ODMR technique, and can be paired with more sophisticated quantum sensing protocols, such as nuclear magnetic resonance (NMR) \cite{holzgrafe2020nanoscale} or spin electron double resonance (SEDOR) \cite{knowles2014observing}. Techniques such as optical tweezers \cite{Geiselmann2013} and surface chemical functionalisation \cite{jung2021surface} can be combined with our dual-modal approach for precise localisation of nanodiamonds with respect to subcellular organelles, such as mitochondria \cite{Di2021}. The targeted delivery of nanodiamonds to subcellular regions would enable probing of potential hot spots, correlating local biochemical events with thermogenesis. This could be used to address the topic of nanoscale temperature gradients in live cells \cite{Macherel2021} and be further extended to studying non-biological soft matter. Active rheology techniques offer an opportunity to further explore the relationship between external forces and the spring constant in cells. Combining nanodiamond sensors with super-resolution imaging techniques \cite{rittweger2009sted} in a multi-scale imaging setting offers an exciting opportunity to probe physical properties in the context of their biological environment.

\bmhead{Acknowledgments}
We would like to thank Ljiljana Fruk,  David Jordan, Thomas Krueger, Erik Miska, Brian Patton, Hannah Stern, Ross Waller and Hengyun Zhou for insightful discussions. This work is supported by the Gordon and Betty Moore Foundation Grant (GBMF7872). Q.G. acknowledges financial support by the China Scholarship Council, the Cambridge Commonwealth, European \& International Trust and the Pump Priming Grant from the Cambridge Centre for Physical Biology. L.S. acknowledges support from the Winton Programme for Sustainability and Robert Gardiner Memorial Scholarship. S.B. acknowledges financial support from EPSRC (PhD Studentship EP/R513180/1), the Alireza Studentship of Lucy Cavendish College and the German Academic Scholarship Foundation. N.S. acknowledges support from the Sperling Studentship. H.S.K. acknowledges the Royal Society University Research Fellowship.



\section*{Methods}
\subsection*{Sample preparation}\label{sec1}
\subsubsection*{Sensing chip}

The resistivity of gold at temperature $T$, $R(T)$, is linear in the temperature range we probe and the measured RTD resistance is converted into temperature using,
\begin{equation}
    T=T_0+(R(T)/R(T_0)-1)/\eta,
\end{equation}
where $T_0$ is the reference temperature at which $R(T_0)$ is measured. $T_0$ is measured with a thermocouple before each chip is used to calibrate the sensor. An experimentally determined fitting constant $\eta=2.44(12)\times10^{-3}\,\mathrm{/^\circ C}$ is used in the conversion.
A polydimethylsiloxane (PDMS) open-top well (with an approximate volume of 400 $\mu$L) is plasma-bonded on top of the substrate when performing experiments involving fluids.

\subsubsection*{Glycerol and GCX suspension}
To make a suspension of nanodiamonds in glycerol, \SI{100}{\micro\liter} of \SI{1}{\milli\gram\per\milli\liter} \SI{50}{\nano\meter} carboxylic-acid terminated nanodiamonds (FND Biotech, Taiwan) in water were centrifuged for 5 minutes at 12000 g. Excess water was removed and the nanodiamond pellet was resuspended in \SI{10}{\milli\liter} of $\ge 99\%$ glycerol (Sigma Aldrich, UK). 

To make the glycerol-crosslinked xanthan (GCX), we transfer \SI{9.9}{\gram} of glycerol-nanodiamond suspension prepared as mentioned above into a beaker and begin agitating by using a magnetic stirrer at a rate of 100 RPM. Then \SI{0.1}{\gram} of xanthan gum powder (Sigma Aldrich, UK) is slowly added to the glycerol. Once all the xanthan is fully incorporated into the glycerol, the beaker is heated to 80°C with the stirring rate increased to 1000 RPM for 1 hour to facilitate crosslinking.

\subsubsection*{Cell preparation}
We use the cervical cancer cell line HeLa as a model system for studying intracellular temperature changes. Prior to the experiment, HeLa cells (ATCC CCL-2) are incubated in an incubator at \SI{37}{\celsius} with $5\%$ \ce{CO_2}. To ensure cells adhere to the sensing chip, the substrate is coated in Geltrex basement membrane product (ThermoFisher Scientific, UK). Sufficient cell coverage was achieved by inoculating 10 $\mu$L of 1x10$^6$ cells mL$^{-1}$ in 390 $\mu$L Dulbecco's modified eagle medium (DMEM) supplemented with 0.11 g L$^{-1}$ sodium pyruvate and 10\% fetal bovine serum (FBS) directly into the PDMS well. Cells were left to adhere to the substrate for a minimum of 12 hours prior to the addition of nanodiamonds.

To ensure an uptake of $\sim 2$ nanodiamonds per cell, 1 $\mu$L of the nanodiamond stock solution was inoculated with the cells in the well for a minimum of 4 hours. Excess nanodiamonds were washed off with phosphate buffered saline (PBS). To compensate for the reduction in CO$_2$ concentration during the ODMR experiments, the cells were innoculated in Leibovitz's L-15 media (ThermoFisher Scientific, UK) supplemented with 10\% FBS for all experiments.

To compensate for the temperature rise due to microwave heating, ODMR experiments were performed in an incubator (Digitalpixel, UK) maintained at $33^\circ \mathrm{C}$, which was given more 12 hours prior to the experiments to thermalise.

\subsection*{Uptake verification}
We take particular caution in verifying the nanodiamonds are internalised inside the cells, to avoid measuring nanodiamonds attached to the surface of the cell. To achieve this, a series of 3D confocal scanning images on a Leica SP5 are taken to establish that the nanodiamonds are situated among the mitochondrial network which has been dyed with 100 nM MitoTracker Green FM (Sigma Aldrich, UK) (Supplementary Material Section 9). The observation that there are mitochondria above, below and surrounding the nanodiamond observed confirm that the nanodiamond is internalised.

\subsection*{Tracking system}
We use a home-built confocal for interrogating the nanodiamond fluorescence. The collimation lens on each collection arm is defocused in opposite directions so that one arm collects at a higher plane and the other at lower plane than the laser focus. This is used for fast particle tracking in the longitudinal direction.
Experimentally, we use a custom Teansy 4.0 micro-controller (MCU) for fast-feedback control. The photon counts are collected via a high-bandwidth counter which are read out by the MCU. The fitting algorithm (implemented in Arduino IDE) uses a linear search and quadratic minimization algorithm. All computation is performed within $\SI{6}{\micro \second}$.

The collection optics are separately aligned and the collimation lenses are displaced in opposite directions so that the collected PL is $70\%$ of the maximum.

\subsubsection*{Static tracking accuracy}
The inferred trajectory of a stationary particle still appears moving. This results from intrinsic tracking error, set-up drift and other sources of movements. These noises are quantified by tracking a stationary nanodiamond (Supplementary Material Section 8). 

\subsubsection*{Dynamic tracking accuracy}
Before the output voltage of the MCU control unit is applied to the actuators, it is additively combined with an external voltage, which can either be \SI{0}{\volt} or a user-generated signal using an external signal source. The latter is used for characterising the dynamic tracking performance. We generate voltages mimicking a Brownian motion particle and compare the extracted trajectory with the known trajectory to quantify dynamics tracking accuracy.

\subsection*{ODMR}
For a 200-point sampling MW frequency ODMR spectrum, a full ODMR scan is completed in \SI{2}{\milli\second}. The scans are acquired for \SI{160}{\milli\second}  (80 full scans) within a \SI{200}{\milli\second} duty cycle. The APD counters are gated such that they do not collect APD counts during the \SI{40}{\milli\second} off time. The ODMR uses a global synchronisation clock with \SI{100}{\kilo\hertz} frequency and gathers one PL reading on each clock edge.

During the \SI{40}{\milli\second} off time, the heater is switched on using a solid-state switch for \SI{30}{\milli\second} with \SI{5}{\milli\second} buffer time before and after. The solid-state switch has a switching rise and fall time of approximately \SI{0.5}{\milli\second}. This avoids any magnetic field interfering with ODMR due the current supplied to the on-chip heater. 

\subsection*{Data processing}
\subsubsection*{ODMR data processing}
The number of photon counts acquired from both APD1 and APD2 (shown in Supplementary Material Fig. S1) within \SI{160}{\milli\second} is read out ($2\times200\times80$ values, for two APDs, with 200 frequency points taken sequentially, and then 80 full repeated scans), converted to counts per second (by dividing by \SI{10}{\micro\second}), summed to give a total counts from both APDs, and stored as $80\times200$ values. Subsequently, each 80 rows of data (i.e. all data gathered within \SI{160}{\milli\second} time frame) is averaged (giving 200 values in counts per second) and stored as intermediate results for post-processing. Subsequent data processing does not use the finely timed raw data due to the long processing time. For a typical emitter giving \SI{1}{\mega\cps} emission, in \SI{10}{\micro\second} read out time, each APD reads about 5 photon counts.

\subsection*{ODMR data fitting methods}
All data presented in this work uses the interpolation method (Supplementary Material Section 4). This method fits the entire dataset using piece-wise linear interpolation to define an interpolation function and subsequently fits subsections of the data with this interpolation function to find any shifts in frequency, $\delta f$. These subsections are defined to be independent bins of \SI{400}{\milli\second} ODMR data and result in a sequence of frequency shifts. We then average consecutive $n_f$ frequency shifts to get a better estimate of the frequency shift. The error in frequency shift is estimated using the standard error of the $n_f$ frequency shifts considered.  
The resultant shift in frequency is converted into a temperature shift presented in the main text. See Supplementary Material Section 4 for more details and motivation.

\subsubsection*{Computation of MSD and associated errors}
We gather a sequence of 3D locations of the nanoparticle, denoted by $\mathbf{r}(n)=(x(n), y(n), z(n))$, where $n=1,2,3,\cdots, N$ are the time indices. To compute the 1D time-averaged mean square displacement in the $x$ direction, we compute the time average of the square of the pair-wise differences, $\xi^x_\tau(i) = x(i+\tau)-x(i)$, by
\begin{equation}
\mathrm{MSD}_x(\tau) = \frac{1}{K} \sum_{i=1}^{K} {\xi^x_\tau}^2(i),
\end{equation}
where $K=N-\tau$. As each $x$ is a stochastic variable, the $\mathrm{MSD}$ is a stochastic variable with inherent variance, irrespective of measurement noise. To quantify this inherent stochastic variance, an estimator is needed based on the measured values of locations, considering the intrinsic correlation between $\xi^x_\tau(i)$ and $\xi^x_\tau(j)$ when $\lvert i-j\rvert <\tau$ as they are based on the same underlying trajectory data. We follow the approach in Kim et al. \cite{kim2015quantification}, in the limit $\tau \ll K$ and $K\gg1$, where the variance of $\mathrm{MSD}_x(\tau)$ is estimated by,
\begin{equation}
\langle \varepsilon_{\mathrm{MSD}_x} ^2(\tau) \rangle = \frac{4}{K} \sum_{i=1}^{\tau} \left(\frac{1}{K-i} \sum_{\alpha=1}^{K-i} \left[ \xi^x_\tau(i+\alpha)\xi^x_\tau(\alpha)\right]^2\right).
\end{equation}
Since there is no correlation between the three spatial directions, 
\begin{equation}
\langle \varepsilon_{\mathrm{MSD}_{xy}} ^2(\tau) \rangle = \langle \varepsilon_{\mathrm{MSD}_x} ^2(\tau) \rangle +\langle \varepsilon_{\mathrm{MSD}_y} ^2(\tau) \rangle.
\end{equation}
The errorbar in Fig. 3 \textbf{(b)} in the main text is computed using this method. It is estimated that in the limit where localisation error is small compared to the stochastic variation due to Brownian motion, the error in determining $D$ scales with $\sqrt{\tau/K}$. The exact proportionality constant depends on the specific method used. 

We estimate the system noise floor to be at $10^{-4}\,\mathrm{\mu m^2}$, limited by the precision of the galvo mirror and signal sources used. The combined error is given by the noise floor if the statistical error derived above or the MSD itself is less than the noise floor.

\subsubsection*{Computation of complex modulus}
At thermal equilibrium the complex modulus, $G^*(f)=G^\prime(f)+iG^{\prime\prime}(f)=\vert G^*(f)\vert e^{i\delta (f)}$, is related to the MSD by,
\begin{equation}
    \tilde{G}(s) = \frac{k_\mathrm{B}T}{\pi r s \langle \Delta \tilde{r}^2(s)\rangle},
\end{equation}
where $\tilde{G}(s)$ and $\langle \Delta \tilde{r}^2(s)\rangle$ are the Laplace transform of the complex modulus, $f = 1/\tau$ and the MSD with respect to the Laplace variable $s$. Using the method developed by Mason, \cite{Mason2000} this is related to the frequency dependence of $G$ by,
\begin{equation}
    \vert G^*(f)\vert \approx \frac{k_\mathrm{B}T}{\pi r \langle \Delta r^2(1/f)\rangle\Gamma(1+\alpha(1/f))},
\end{equation}
where $\Gamma$ is the $\Gamma$-function. The loss tangent, $\delta$, follows,
\begin{equation}
    \delta\left(\frac{1}{f}\right) = \frac{\pi}{2}\alpha\left(\frac{1}{f}\right).
\end{equation}
Here $\alpha(1/f)=\frac{d\ln{\mathrm{MSD}}}{d\ln{\tau}}\vert_{1/f}$ is the gradient of the MSD curve on a log-log plot. 

\subsubsection*{Power spectral density}
The PSD, $\langle {\Delta x}^2(f) \rangle$ of motion in one axis was calculated using Welch’s method (welch function in scipy.signal python module), averaged over 28.8 s. The total PSD is given by the sum of PSD in the two transverse axes, $\mathrm{PSD} = \langle {\Delta x}^2(f) \rangle + \langle {\Delta y}^2(f) \rangle$. PSD data presented in the maintext is taken at $f=40\,\mathrm{Hz}$.




\setcounter{figure}{0}
\renewcommand{\figurename}{Fig.}
\renewcommand{\thefigure}{S\arabic{figure}}

\setcounter{table}{0}
\renewcommand{\tablename}{Table}
\renewcommand{\thetable}{S\arabic{table}}

\setcounter{section}{0}

\onecolumn{
\section*{\begin{center}
    \textbf{Supplementary Information for \\
`Simultaneous nanorheometry and nanothermometry using intracellular diamond quantum sensors'}
\end{center}}

\section{Confocal microscope set-up}

\begin{figure*}[b]
    \centering
    \includegraphics[width=\textwidth]{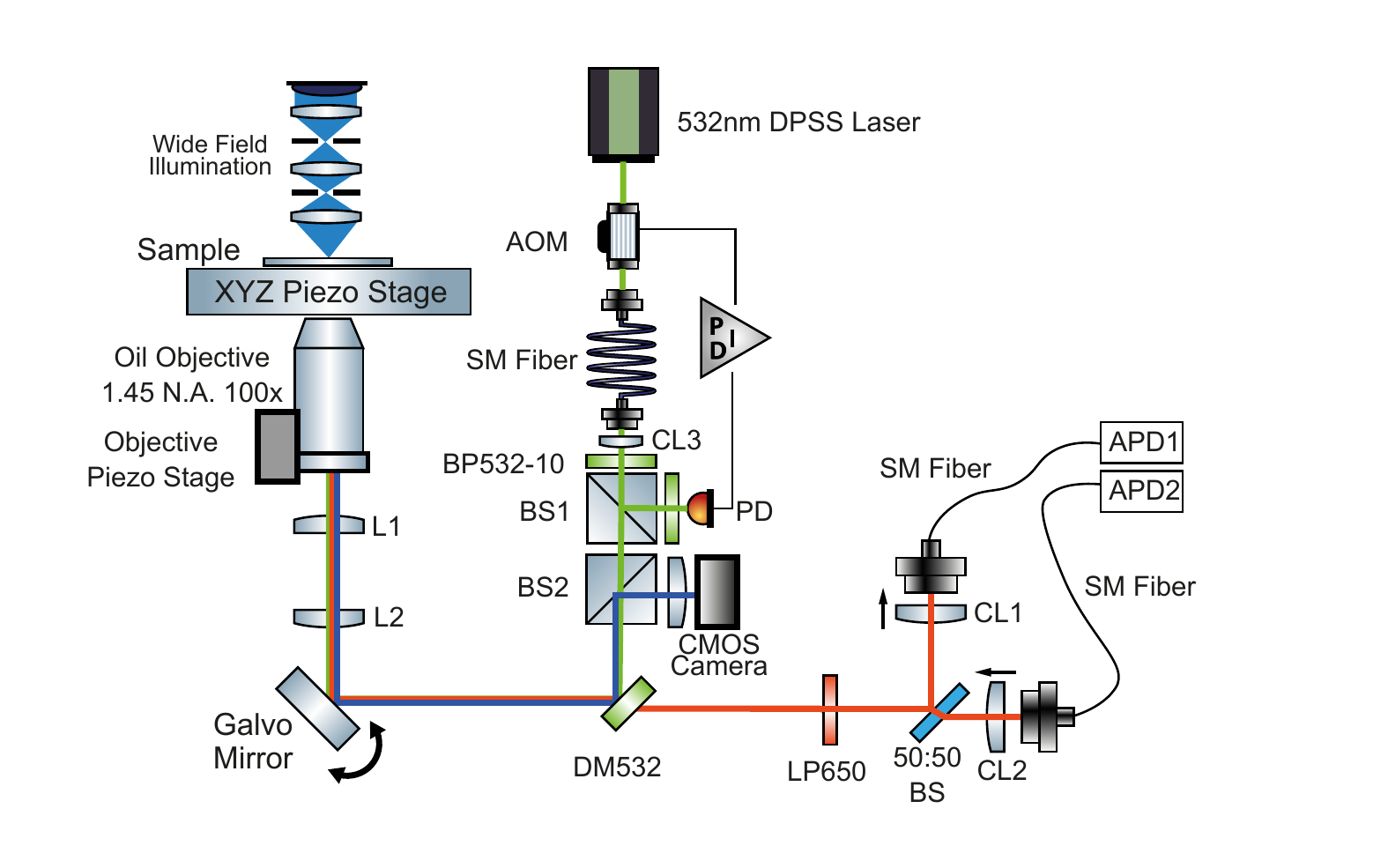}
    \caption{Optical setup. A confocal set-up is used to excite the NVs with \SI{532}{\nano\meter} and their fluorescence is collected between \SI{650}{\nano\meter} and \SI{800}{\nano\meter}. A 2D galvonomic mirror with a Keplerian telescope are used for xy positioning. The objective is mounted on a piezo-stage. The 2D galvonomic mirror and piezo-stage are used for single particle tracking. L1/L2: achromatic lens. BS1/BS2: nonpolarising beamsplitters. CL1/CL2/CL3: collimation lens. APD: Avalanche photodiode. SM Fiber: Single-mode fiber. AOM: Acousto-optic modulator. DM: Dichroic mirror. LP: Long-pass filter. BP: Band-pass filter.}
    \label{fig:ConfocalSetup}
\end{figure*}

The nitrogen-vacancy centres (NVs) are optically read out with a laser scanning confocal microscopy setup to achieve high spatial resolution. The confocal imaging part of the setup (Fig. \ref{fig:ConfocalSetup}) consists of an excitation arm and a collection arm. A \SI{532}{\nano \meter} laser (Ventus 532, Novanta Photonics) provides the optical excitation. The laser  is collimated after the single-mode fibre and overfills the back aperture of an oil immersion objective (Nikon CFI Plan Apo Lambda 100X Oil MRD01905, 1.45 N.A.). After collimation, $10\%$ of the beam is split off to a photodiode (Si switchable gain detector, Thorlabs) for power monitoring. This power measurement is also used to stabilise the laser power via a high-bandwidth (\SI{100}{\kilo \hertz}) PID controller (SIM900, Stanford Research Systems) which feeds back to a MT80-A1.5-VIS Acousto-optic modulator, Opto-Electronic.\\

The fluorescence from the NVs is collected through the objective and passes along the collection path where it is filtered by a \SI{532}{\nano\meter} dichroic mirror, and a \SI{650}{\nano\meter} long-pass collection filter, split into two collection pathways equally via a 50:50 plate beamsplitter, each focused via an achromatic lens into another single-mode fibre. Two avalanche photodiodes (APDs, SPCM-AQRH-14-FC, Excelitas technologies) are used for photon detection. The output pulses are counted by two DAQ cards (PCIe6321, PCIe6323, National Instruments) or by two custom high-bandwidth counters for the tracking system. All counter readings are synchronised to a global \SI{100}{\kilo \hertz} clock, generated by a DAQ card (PCIe6321, National Instruments).\\
The collimation lenses on each collection arm are defocused in opposite directions to 70\% of the in-plane counts. This results in one arm collecting counts $~50\ \mathrm{nm}$ above and the other $~50\ \mathrm{nm}$ below the laser focus. This is used for fast particle tracking in the longitudinal direction. The digital computation takes $<\mathrm{6\,\mu s}$ to complete and thus the feedback can potentially be reduced to sub-\SI{10}{\micro\second} for faster moving particles if other parts of the system permit.
A two-dimensional galvanometer mirror (GVS002, Thorlabs, UK) and an objective piezostage (DRV517, Thorlabs, UK) are used for feedback-based single particle tracking. The sample is mounted on an XYZ piezostage (NPoint, Inc., US) which provides a much larger range than the galvonomic mirrors and is used for slow sample positioning.

\section{Diffusion constant statistics based on imaging experiments}
To verify that our tracking method does not significantly overlook the fast diffusing particles, we used a commercial scanning confocal microscope to build up the statistics for diffusion constants. The tracking limit $D=5\times10^4\,\mathrm{nm^2/s}$ covers 99\% (Fig. \ref{fig:ComparisonTrackingPerformance}) of all particles observed in a commerical confocal microscope.

 \begin{table*}[h]
     \centering
     \resizebox{\textwidth}{!}{
     \begin{tabular}{|c|c|c|c|c|}
          \hline
          Nanoparticle & Diameter & $D_{\mathrm{max}}$ (\SI{}{\square{\micro\meter}\per \second}) & Cell/Organism & Ref.\\
          \hline \hline 
          Polyplex & \SI{266}{\nano\meter} & 0.0032 & HuH-7 (human hepatome cells) & Dupont et al\cite{dupont2013three}\\
          Polystyrene & \SI{110}{\nano\meter} & 0.0003 & Opossum kidney (OK) proximal tubule cells  & Lanzano et al\cite{lanzano2014orbital}\\
          Quantum dot & - & 0.03 &  Human lung carcinoma A549 cells  & Li et al\cite{li2018intracellular}\\
          \hline
     \end{tabular}
     }
     \caption{Diffusion speed of nanoparticles in other cells of human origin. The maximum speed tracked in this work (\SI{0.05}{\square{\micro\meter}\per\second}) exceeds all use case of single-particle tracking in individual cells.}
     \label{tab:TrackingPerformance}
 \end{table*}
 
 \begin{figure*}[h]
     \centering
     \includegraphics[width=\textwidth/2]{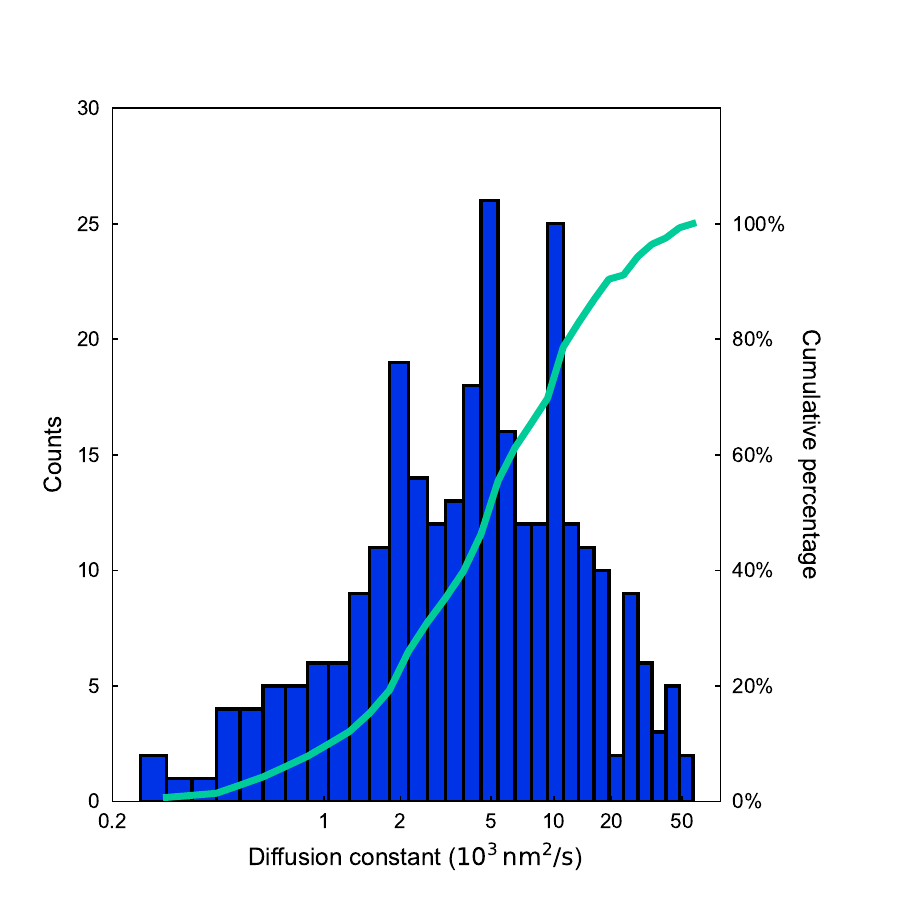}
     \caption{Comparison between tracking performance and intracellular tracking requirement. Distribution of diffusion constants for nanodiamonds in HeLa cells extracted from image analysis based on confocal fluorescence images. The maximum tracking speed of our system, $D=5\times 10^4\,\mathrm{nm^2/s}$ covers 99\% of all nanodiamonds observed, thus not inducing a bias. }
     \label{fig:ComparisonTrackingPerformance}
 \end{figure*}

\section{Double-plane orbital tracking}

\begin{figure*}
    \centering
    \includegraphics[height = 5cm]{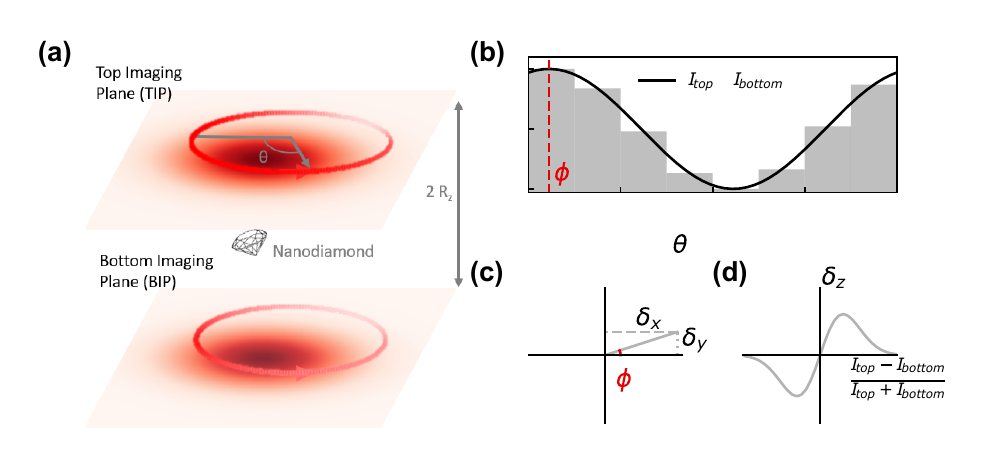}
    \caption{\textbf{(a)} Orbital Tracking. Photons are collected from two imaging planes which are separated by $2\,R_z$. The tracker orbits the last known position of the nanodiamond. \textbf{(b)} and \textbf{(c)} The correction in the xy plane, $\Delta_x$ and $\Delta_y$, are calculated by fitting the collected data (grey) to a sine wave (black).  calculating the amplitude, $\delta \propto \sqrt{\delta_x^2 + \delta_y^2}$, and direction, $\phi$, of the correction. \textbf{(d)} The axial correction, $\Delta_z$, is calculated by finding the difference in counts between the top imaging plane and bottom imaging plane. }
    \label{fig:Orbital Tracking}
\end{figure*}

In the tracking process, the scanning mirrors are actuated to undergo a circular motion with radius $R_{xy}$ and period $T$ such that the deflection of each mirror at time t, can be expressed as
\begin{align}
    x &= R_{xy}\cos{\theta}\\
    y &= R_{xy}\sin{\theta}
\end{align}

\noindent where $\theta = \frac{2\pi t}{T}$. The axial collection planes are displaced so that photons are collected from $z=\pm R_z$ simultaneously. The signals collected are binned into eight bins each corresponding to $\frac{1}{8}$ of the orbit. Thus, there are eight points per plane and two planes in total, giving 16 data points. These are at locations $(R_{xy}\cos{(\theta_n)}, R_{xy}\sin{(\theta_n)}, \pm R_z)$ where $\theta_n$ is the continuous angle $[n\pi/4, (n+1)\pi/4]$ for $n=0, 1, \cdots, 7$. We denote the corresponding photo luminescence (PL) as, $I_{z, n}$ where $z$ = top denotes the top orbit and $z$ = bottom denotes the bottom.\\
As the confocal volume addresses hundreds of NVs, each of which is assumed to be randomly oriented, the point spread function (PSF) is assumed to be unaffected by the emission dipole orientation of the NV and of a Gaussian form, with $w_{xy}$ and $w_z$ being the $1/e^2$ radius of the PSF in the radial and the axial directions. For the confocal collection volume centered at $(x, y, z)$ which orbits around a fixed point $(0, 0, 0)$ and a nanodiamond placed at $(\delta_x, \delta_y, \delta_z)$, the collected photon counts are,

\begin{equation}
    I(\delta_x, \delta_y, \delta_z, x,y,z) =
    I_{\mathrm{top/bottom, C}} R(\delta_x, \delta_y,x,y)Z(\delta_z,z),
\end{equation}

\noindent where $I_{\mathrm{top/bottom, C}}$ is the count rate on each collection arm when the emitter is at the point about which the confocal volume is orbiting and the radial and axial parts are,
\begin{align}
    & R(\delta_x, \delta_y, x,y)  = \exp{\left( - 2\frac{(x-\delta_x)^2 + (y-\delta_y)^2}{w_{xy}^2} \right)}, \\
    & Z(\delta_z, z)  = \exp{\left( - 2\frac{(z-\delta_z)^2}{w_{z}^2} \right)}.
\end{align}

\noindent When the nanodiamond deviates from the focal spot by a small amount, $\delta_{x, y} \ll w_{xy}$, 

\begin{equation}
R(\delta_x, \delta_y, x, y) \approx \exp{\left(-2\frac{x^2+y^2}{w_{xy}^2} \right)}\left[1+\frac{4x}{w_{xy}^2}\delta_x+\frac{4y}{w_{xy}^2}\delta_y\right]
\end{equation}

\noindent Thus, along the circular orbit, the summed PL for the top and bottom axial planes is 
\begin{equation}
    I_n = I_{\mathrm{top}, n}+I_{\mathrm{bottom}, n} \approx I'\left[ 1+\delta\cos{(\theta_n + \phi)}\right],
\end{equation}
where $I' = \left[I_{\mathrm{top, C
}}Z(\delta_z, R_z)+I_{\mathrm{bottom, C}}Z(\delta_z, -R_z)\right]\exp{\left(-2\frac{R_{xy}^2}{w_{xy}^2} \right)}$, $\delta^2=(\delta_x^2+\delta_y^2)/\varepsilon_{xy}^2$, $\varepsilon_{xy}=\frac{w_{xy}^2}{4R_{xy}}$ and $\tan{\phi}=\delta_y/\delta_x$. The parameters $I'$, $\delta$ and $\phi$ are fitted using a least square fitting algorithm. An illustration of this process is show in Fig. \ref{fig:Orbital Tracking}.\\
As there are two collection planes going into two detectors, there is a slight mismatch between the collection readout due to differences in the fiber coupling efficiency, an unbalanced beam splitting ratio or misalignment in the optics. As a result $I_{\mathrm{top, C}} \neq I_{\mathrm{bottom,C}}$ and we require an additional term $G=\frac{I_{\mathrm{bottom, C}}-I_{\mathrm{top, C}}}{I_{\mathrm{bottom, C}}+I_{\mathrm{top, C}}}$ which is a setup dependent experimentally determined parameter. To compute the feedback parameter for the axial direction, the total PL from the top and bottom planes are $I_{\mathrm{top}, \mathrm{total}} = \sum_nI_{\mathrm{top}, n}$ and $I_{\mathrm{bottom}, \mathrm{total}} = \sum_nI_{\mathrm{bottom}, n}$. Then we compute the ratio,
\begin{equation}
    r\equiv\frac{I_{\mathrm{bottom}, \mathrm{total}}-I_{\mathrm{top}, \mathrm{total}}}{I_{\mathrm{bottom}, \mathrm{total}}+I_{\mathrm{top}, \mathrm{total}}} \approx \frac{G-\delta_z/\varepsilon_z}{1-G\delta_z/\varepsilon_z},
\end{equation}
where $\varepsilon_z=\frac{w_z^2}{4R_z}$.\\
After obtaining the parameters, $\delta$, $\phi$ and $r$, we compute the corrections by,

\begin{equation}
    (\Delta_x, \Delta_y, \Delta_z) = \Bigg(\delta \varepsilon_{xy}\cos{\phi}, \delta\varepsilon_{xy}\sin{\phi}, \frac{r-G}{rG-1}\varepsilon_z\Bigg)
\end{equation}
where $\Delta_{x,y,z}$ are the estimates of $\delta_{x,y,z}$ respectively. The parameters used are listed in Table \ref{tab:orbitalTrackingParameters}. Note that the XY parameters are real experimental values.

\begin{table}[]
    \centering
    \begin{tabular}{|c|c|}
        \hline
        Parameter & Value \\
        \hline \hline
        T & \SI{9.6}{\milli\second} \\
        \hline
        $R_{xy}$ & \SI{50}{\nano\meter}  \\
        \hline
        $w_{xy}$ & \SI{260}{\nano\meter} \\
        \hline
        $R_{z}$ & \SI{200}{\nano\meter} \\
        \hline
        $w_{z}$ & \SI{200}{\nano\meter} \\
        \hline
        $G$ & $[-0.2, +0.2]$ \\
        \hline
    \end{tabular}
    \caption{Parameter used for feedback localisation locking. }
    \label{tab:orbitalTrackingParameters}
\end{table}

\section{Interpolation Function}\label{section:interpolation}
Several lineshapes have been used to fit ODMR data including a double Lorentzian \cite{sotoma2021situ} and a single Lorentzian \cite{simpson2017non}. The double Lorentzian is a 7 parameter function and the single Lorentzian is a 4 parameter function. For an ODMR where the shape of the lineshape does not change, both of these fitting functions can be adjusted to have only 2 free parameters, the central frequency, $f_0$ and the counts, $\Lambda_0$. For ensembles of NVs the ODMR lineshape is a combination of the spectra from the individual NVs, each with their own central frequency and strain. As a result, the ODMR from ensemble NVs shows inhomogeneous spectral broadening and is not necessarily best described by a standard function.\\
We introduce the interpolation function as an alternative lineshape which can be used to fit ODMR data with two fitting parameters. For each pair of adjacent data points $(f_1, \mathrm{PL_1})$ and $(f_2,\mathrm{PL_2})$ the linear interpolation function is defined by,

\begin{equation}
L^{\mathrm{Interp}}(f)\rvert_{f_1}^{f_2}  =  \frac{1}{\mathrm{\Lambda_0}}\Biggl(\mathrm{PL}_1  +  (f - f_1)  \frac{(\mathrm{PL}_2 - \mathrm{PL}_1) }{ (f_2 - f_1)}\Biggr).
\end{equation}\\
The lineshape function is then fitted to the data using

\begin{equation}
\mathrm{PL}(f) = \mathrm{\Lambda_0} L^{\mathrm{Interp}}(f - \delta f), 
\label{interpolation}
\end{equation}\\
where the two fitting parameters are the off resonance PL, $\mathrm{\Lambda_0}$, and the shift in frequency, $\delta f$.\\
To define the interpolation function we use the average over the full length of the ODMR dataset. This function was used to fit $400\,\mathrm{ms}$ of data and the offset in frequency was extracted. These frequency offsets are then further averaged so that each data point represents $12\,\mathrm{s}$ of data. The frequency offset is converted to temperature using the calibrated value for $\kappa$.

 \subsection{Comparison of interpolation function against other fitting methods}

 \begin{figure*}[h]
    \centering
    \includegraphics[width=\textwidth]{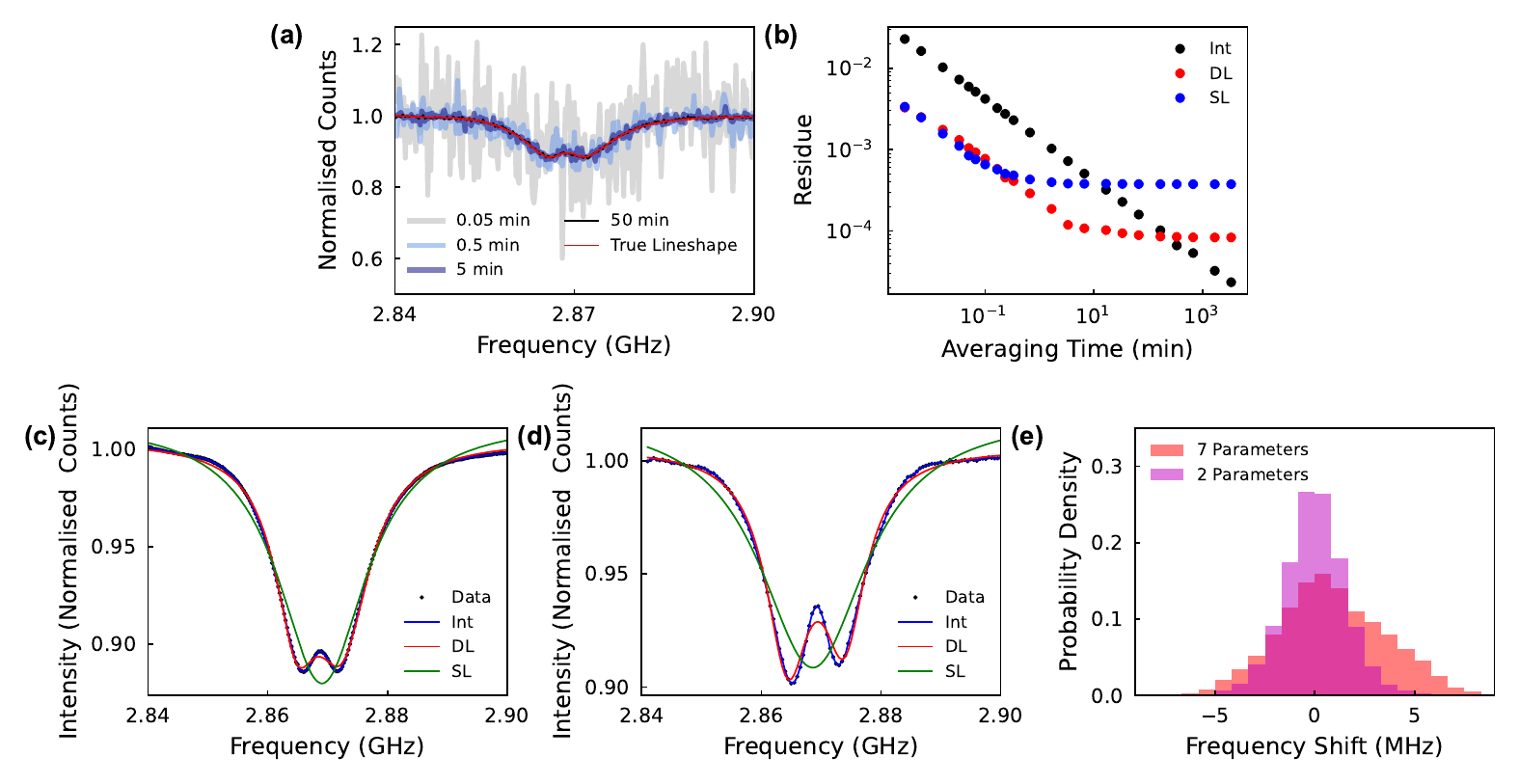}
    \caption{Comparison of fitting lineshapes. \textbf{(a)} The noise in the interpolation function for different lengths of datasets. \textbf{(b)} The average residue in the normalised fit for the Interpolation function (blue), double-Lorentzian (red) and single-Lorentzian (green) for different lengths of dataset. \textbf{(c)} and \textbf{(d)} Finding the best lineshape for the dataset. Fitting the data (black) with a Interpolation function (Int - blue), double-Lorentzian (DL - red) and single-Lorentzian (SL - green). \textbf{(e)} Probability of achieving a frequency shift in the central frequency for a generated set of data with Gaussian noise at constant temperature. The double-Lorentzian fit was made with both 7 parameters (red) and 2 parameters (pink). The fixed parameters were extracted from the fitting in (c)}
    \label{fig:Linshapecomparison}
\end{figure*}

The interpolation function offers an advantage as it can be used to fit all lineshapes. In bulk NVs the intrinsic strain is usually quite low. However, in commercially available nanodiamonds there can be large variations in strain which can result in the presence of side peaks and/or assymmetry. The interpolation function is robust to the unusual ODMR lineshapes often seen in nanodiamonds. One caveat for using an interpolation function is that the data must be of a suitable length. As shown in Fig. \ref{fig:Linshapecomparison} \textbf{(a)}, when the data is averaged over short periods of time the interpolation function retains noise from the data. The interpolation function is compared to the `true data' which is the underlying lineshape from a nanodiamond before any noise was added to the system. Noise has been added using a gaussian distribution with sigma calculated from the noise present on the nanodiamond. The average residue between the true data and the fitting model across the frequency range is calculated to quantify how well the function fits the data for a dataset of a certain length. As can be seen in Fig. \ref{fig:Linshapecomparison} \textbf{(b)}, at short time intervals the interpolation function is dominated by noise but for longer time intervals the interpolation function more closely matches the true lineshape than the single Lorentzian $(T>10\,\mathrm{min})$ or the double Lorentzian $(T>100\,\mathrm{min})$.

In order to use a two parameter fitting method, it is important that the fitting function accurately represent the data. This involves both choosing the correct function for the lineshape and making sure any fixed parameters in the function are known. To find the fixed parameters, we average the data over the full dataset. This averaged ODMR is then fit using all fitting parameters (7 for double Lorenzian, 4 for single Lorentzian) and the fixed parameters are extracted. Fig. \ref{fig:Linshapecomparison} \textbf{(c)} shows a comparison for the different fitting models (Interpolation - blue, double Lorentzian - red and single Lorentzian - blue) against the averaged or true data (black). As can be seen in the figure, the intrinsic strain in the nanodiamond results in a splitting between the $m_s = \pm 1$ states and the single Lorentzian struggles to fit this. Fig. \ref{fig:Linshapecomparison} \textbf{(d)} shows an example of an assymetric peak. We also take this opportunity to highlight the improvement in precision by using 2 parameter fitting. For a generated set of data at constant temperature with gaussian noise, the precision of the double Lorentzian 7 and 2 parameter fitting models were compared. As seen in Fig. \ref{fig:Linshapecomparison} \textbf{(e)} the 2 parameter fit has a much narrower spread on the central frequency, $ 1.6\,\mathrm{MHz}$ than the 7 parameter fit, $ 2.7\,\mathrm{MHz}$.\\
As described in section \ref{Sensitivity} the theoretical sensitivity can be calculated for a given lineshape. For the data shown in Fig. \ref{fig:Linshapecomparison} \textbf{(c)}, the theoretical sensitivities were found to be comparable for the interpolation function (\SI{2.1}{\kelvin \per \sqrt{\hertz}}), double-Lorentzian  (\SI{2.2}{\kelvin \per \sqrt{\hertz}}) and single-Lorentzian \SI{2.0}{\kelvin \per \sqrt{\hertz}}. As sensitivities are comparable and in this paper we work with long datasets, due to the robustness of the fitting model we use the interpolation model to fit the data. A step by step description of the fitting model can be seen in Section \ref{section:interpolation}.

\section{Resistance Temperature Detector}
The operation of the Resistance Temperature Detector (RTD) is based on the linearity of electrical resistance of gold over the physiologically relevant temperature range.
The resistance at temperature $T$ is related to the resistance at reference temperature, $T_{\mathrm{ref}}$, as (reproduced from Methods),
\begin{equation}
    R(T)/R(T_{\mathrm{ref}}) = \eta (T-T_{\mathrm{ref}}) + 1,
\end{equation}
where $\eta$ is an experimentally determined constant related to the temperature coefficient of gold and is calibrated using the incubator. The on-chip heater and RTD allow the temperature to be controllably changed. The on-chip heater and RTD allow temperature to be changed by $2\ \mathrm{^{\circ}C}$ within \SI{2}{\minute} and the temperature stability can be maintained within \SI{5}{\milli\kelvin} (data not shown). The positioning of the RTD relative to the CPW and heaters can be seen in Fig. \ref{fig:chip-picture}.

\begin{figure*}[h]
    \centering
    \includegraphics[width=\textwidth]{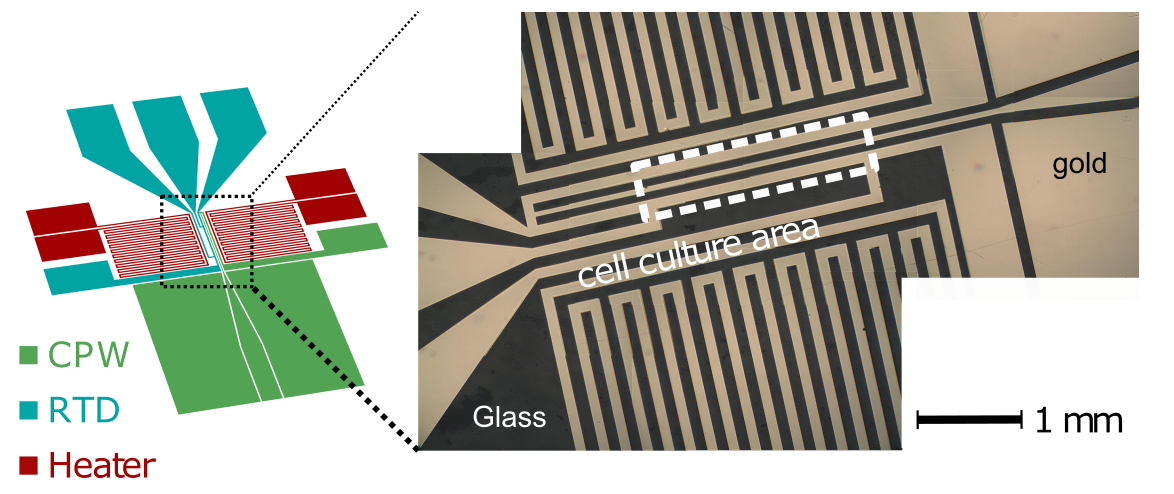}
    \caption{Sensing chip. Left: schematic showing the different components of the pattern. Red: resistive heaters. Cyan: resistive temperature detector (RTD). Green: co-planar waveguide (CPW). In the micrograph on the right, the glass substrate is in black and gold pattern in yellow. The area where we choose to image is labelled with dashed rectangle, labelled ``cell culture area". }
    \label{fig:chip-picture}
\end{figure*}

\section{Experiment values for $\kappa$}
In Table \ref{tab:kappa_table}, we provide a list of NV thermometry experiments reported to-date and the corresponding temperature-frequency conversion coefficients, $\kappa$, used in these work. Due to the differences in ODMR measurement methods and diamond manufacturer, the values vary significantly. When a constant value is assumed from the literature, or from a batch calibration, the actual temperature may be under- or overestimated, thus causing systematic error.
\begin{table*}[h]
    \centering
    \begin{tabular}{|@{\em}p{1.7cm}@{}|@{\em}p{1.4cm}@{}|@{\em}p{6cm}@{}|l|}
    \hline
         $\vert \kappa \vert $ (\SI{}{\kilo\hertz\per\celsius}) & Cellular noise & Comments & Ref.  \\
         \hline \hline
          74 & 30\% & Assumes constant value. Across multiple HeLa cells due to laser induced heating of surface dopamine. & Sotoma et al \cite{sotoma2021situ}  \\
          \hline
          65.4 & 22\% & Mean value of a few calibrated NDs. Temperature rise in multiple \textit{C. elegans} due to chemical treatment with an uncoupler. & Fujiwara et al. \cite{fujiwara2020real}  \\
          \hline
          74 & 8\%  & Assumes constant value. Across one neuron. & Simpson et al \cite{simpson2017non}  \\
          \hline
          \SI{66(11)}{} & - & Investigated $\kappa$ dependence on chemical environment & Sekiguchi et al \cite{sekiguchi2018fluorescent}  \\
          \hline
          \SI{78(12)}{} & - & Mean value of 15 NDs & Yukawa et al \cite{yukawa2020quantum}  \\
          \hline
          77 & - & Assumes constant value & Kucsko et al \cite{kucsko2013nanometre}  \\
         \hline
    \end{tabular}
    \caption{Values used to convert from ODMR central frequency shift to temperature change ($\kappa$). The typical cellular noise is derived from the ratio between the report spread and mean value of the biological process of interest. }
    \label{tab:kappa_table}
\end{table*}

\section{Mathematical limit of ODMR sensitivity}\label{Sensitivity}
After obtaining the ODMR spectra, we use the nonlinear least-square curve fitting method to fit the data to a predefined lineshape function to estimate the shift of the ODMR central frequency. Various lineshape functions have been used previously, such as the double-Lorentzian \cite{sotoma2021situ} typically used in NV sensing based on single NVs and single-Lorentzian \cite{simpson2014vivo} functions. We wish to understand the theoretical limit of parameter estimation based on curve fitting, given the noise we see experimentally. \\
Here, we use the Fisher information to estimate the lower bound on the variance of estimated parameters, known as the Cram\'{e}r-Rao bound (CRB). This is a technique widely used in super-resolution imaging but has not yet been applied in the context of NV thermometry. Given a noise model (Poissonian for shot-noise limited measurements), a fitting function (for example double Lorentzian), and a set of points used to sample the fitting function (for example RF frequencies used to measure ODMR), CRB gives the lower bound on the uncertainty of estimation for any unbiased estimators.\\
To start with, we assume the experimentally measured ODMR spectrum consist of a sequence of measured values of photon counts in a unit time interval, $\mathrm{PL}(f_i)$ taken at various RF frequencies, $f_i$. These are random variables due to experimental noise, and follow the distribution
\begin{equation}
    \mathrm{PL}(f_i) \sim \mathrm{Poisson}(\Lambda_0 L_\theta(f_i)),
\end{equation}
where $\Lambda_0$ is interpreted as the off-resonance PL in a unit time interval and $L_\theta(f_i)$ is the lineshape function (normalised such that $L_\theta(f)=1$ for non-resonant frequencies), parameterised by the constant vector $\theta$. The lineshape function can be the linear interpolation function for which $\theta =(\delta f)$ (a single-element vector) where $\delta f$ is the shift in ODMR central frequency. It can also be the double Lorentzian function in which case $\theta$ is a six-element vector with parameters characterising the contrasts, HWHM's and centres of each constituent Lorentzian. The CRB is applicable to any lineshape function with a variable number of parameters and so the following discussion is kept general.\\
The covariance matrix of the estimated values of $\theta$, denoted by the estimator, $\hat{\theta}$, is given by the Cram\'{e}r-Rao bound as,
\begin{equation}
    \mathrm{cov}(\hat{\theta}) \ge \frac{1}{\Lambda_0}\left[\sum_i\frac{{L'}^2(f_i)}{L(f_i)} \right]^{-1},
\end{equation}\\
where $L'\equiv \partial_{\theta_i}L$ is the gradient of the lineshape function in the $\theta$ space and $L'^2$ is a square matrix with components $(L'^2)_{ij}\equiv \left(\partial_{\theta_i}L\right) \left(\partial_{\theta_j}L\right)$. Here, $i=1,2,3,\cdots,N$ with $N$ being the number of parameters used. The $(\bullet)^{-1}$ should be interpreted as the matrix inverse when $\theta$ has more than one elements or the multiplicative inverse (in the algebra sense) when $\theta$ has only one element. Two square matrices $A$ and $B$ satisfy the relation $A\ge B$ if $\langle i\vert (A-B) \vert j\rangle \ge 0$ for any choice of non-zero vectors $\vert i\rangle$ and $\vert j\rangle$. Specifically, if only the diagonal of the covariance matrix is of interest, then $A\ge B$ implies that every diagonal element of $A$ is greater than or equal to that of $B$. Thus, by choosing the diagonal element corresponding to the central frequency, the CRB gives the lower bound on the uncertainty for frequency estimation. We use this to compute the theoretical sensitivity bound quoted in the main text.

\section{Tracking performance}
\subsection{Stationary tracking performance}
The power spectral density (PSD)(Fig. \ref{fig:stationaryTrackingPerformance}) of the positions of a stationary particle reported by the tracker is computed using \SI{190}{\second} of tracking data in three spatial directions for each PL intensity corresponding to different laser power. In particular, we note that the static tracking noise is limited by the fluorescence (Fig. \ref{fig:stationaryTrackingPerformance} (a) and (b)), such that increasing the fluorescence by using brighter particles or increasing the laser power can improve the tracking performance. A similar effect is shown in the PSD where the static tracking noise is also reduced with increased particle brightness, as shown in all axes in Fig. \ref{fig:stationaryTrackingPerformance}, (c)-(e). 

\begin{figure*}[h]
    \centering
    \includegraphics[width=\textwidth]{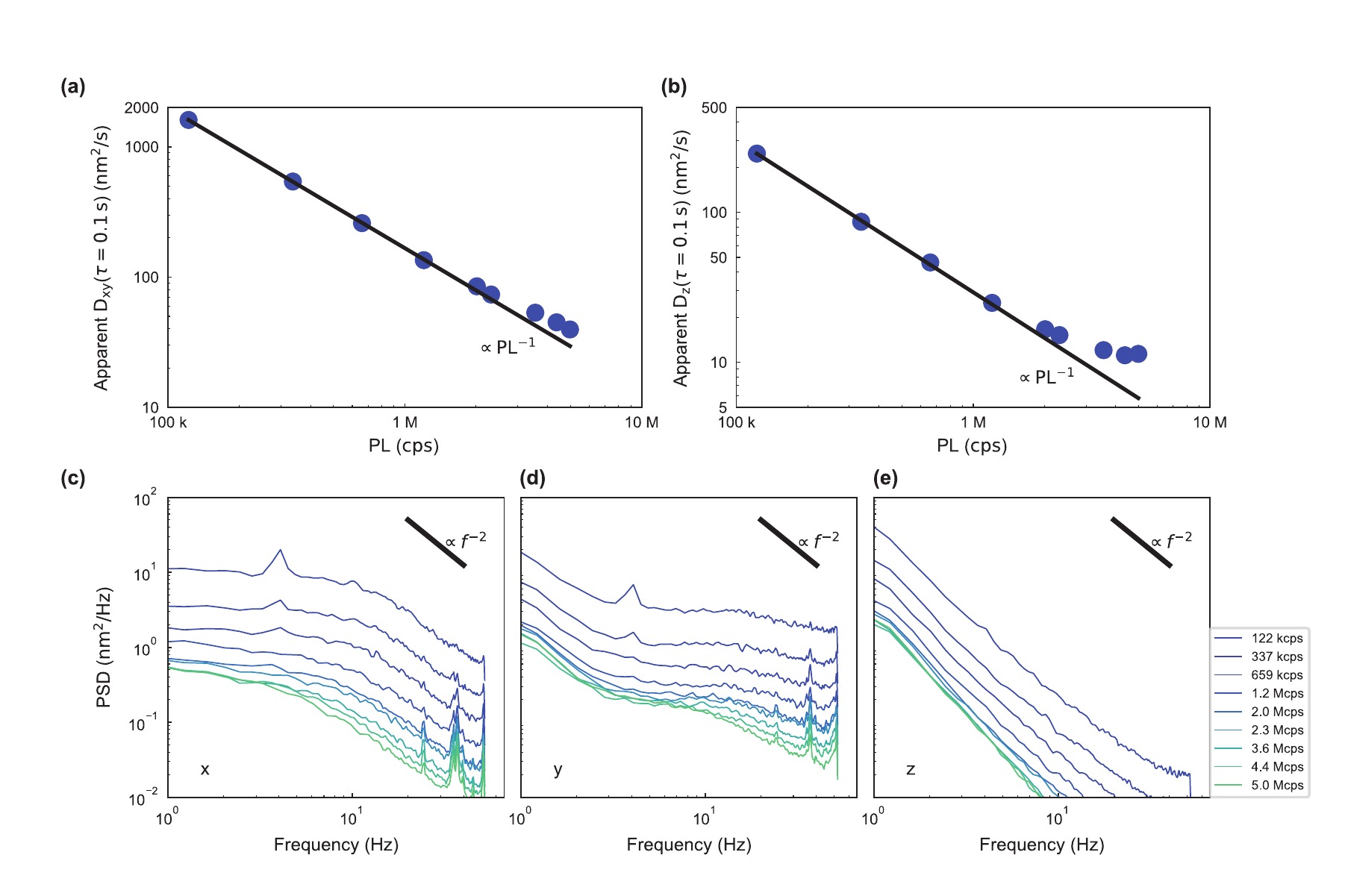}
    \caption{Sensitivity of nanodiamond single particle tracking. \textbf{(a),(b)} The apparent diffusion constants, in XY and in Z direction, of a stationary nanodiamond, limited by photon-shot noise and other noise in the system. The apparent diffusion constant, $D$, scales with $\mathrm{PL}^{-1}$ for small PL emitters. \textbf{(c)-(d)} The power spectral density for individual axis.}
    \label{fig:stationaryTrackingPerformance}
\end{figure*}

\subsection{Dynamic tracking accuracy}
To characterise the dynamic tracking performance, an external, user-generated voltage signal can be additively combined with the output voltage of the micro-controller unit (MCU) before it is applied to the actuators. We use a DAQ card (PCIe 6323, National Instruments) to generate voltages that mimic a 3D random walk. For one dimension, such a random walk is given by,
\begin{equation}
    x(n\tau) = \sum_{i=0}^n \Delta x_i,
\end{equation}
where $\Delta x_i\sim\mathcal{N}(0, 2D\tau)$ is a sequence of randomly generated steps following a Gaussian distribution with a zero mean and a variance of $2D\tau$. $D$ is the diffusion constant input by the user. The time step, $\tau$, is set by the sampling rate of the signal source and is \SI{240}{\micro\second}. A different sequence of random numbers are used for each axis. The timing of the signal generator is again synchronised with the global \SI{100}{\kilo\hertz} clock so that the generated trajectory can be aligned temporally with data obtained from the tracking system or optically detected magnetic resonance (ODMR) experiments.\\
In principle, the signal generator and the tracking system are unaware of each other and thus a nanodiamond fixed on a substrate would appear to be diffusing to the tracking system. The predefined trajectory is then compared with the extracted trajectory from the tracking system to measure the diffusion constant and localisation accuracy (Fig. 2 (e)). As expected the dynamic localisation accuracy is worse at high input diffusion rate, shown in Fig. 2 (f) in the main text.

\section{Uptake studies}
\begin{figure*}[ht]
    \centering
    \includegraphics[width=\textwidth*4/5]{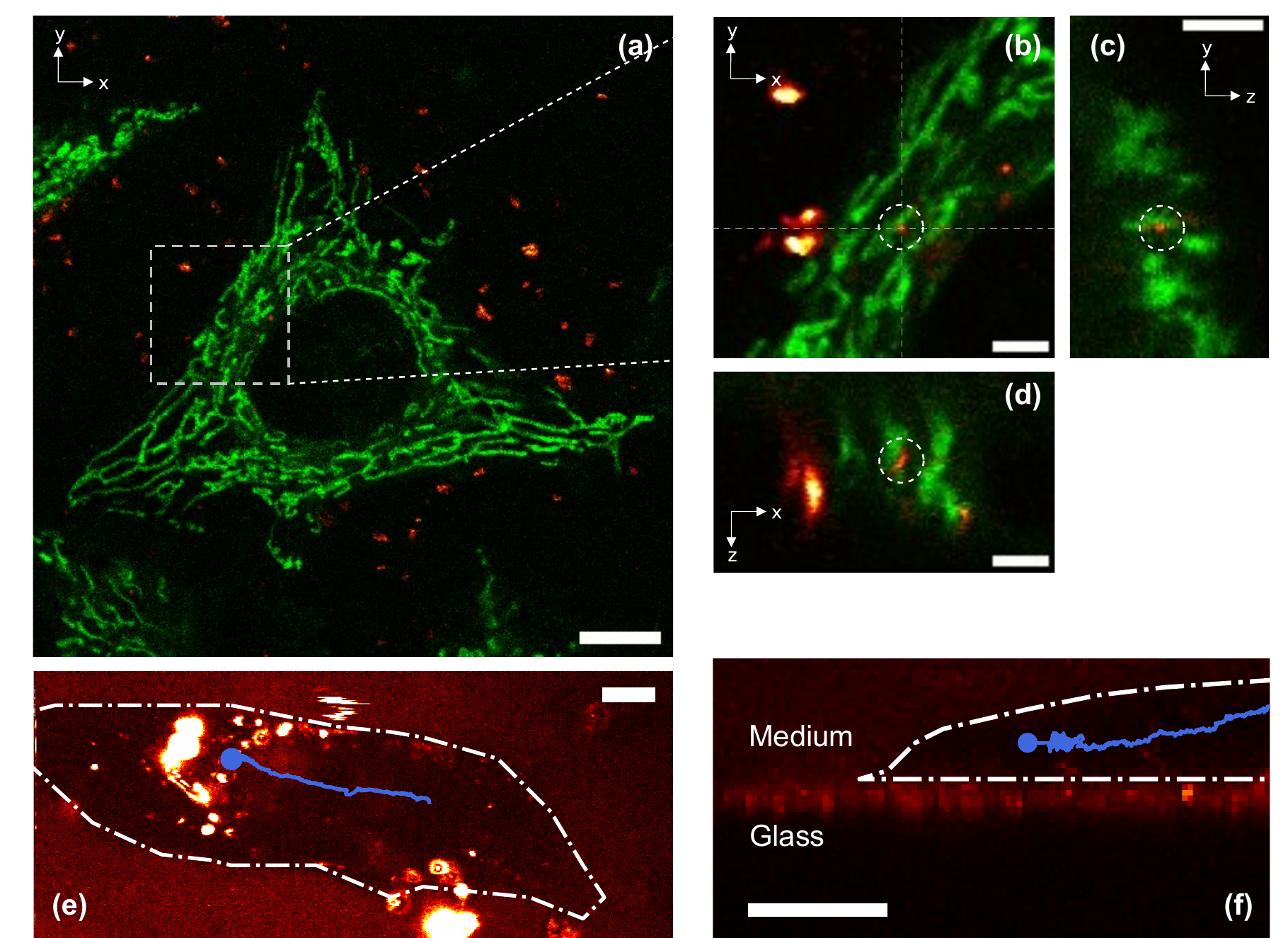}
    \caption{Uptake verification using a laser scanning confocal microscope. \textbf{(a)} Two-colour confocal image showing nanodiamond (red) internalization in a HeLa cell (where the image has been taken 500 nm above the bottom of the cell). Mitochondria (green) are used as a reference for identifying the cell's internal volume. Scalebar is \SI{10}{\micro\meter}. Three-dimensional imaging was performed on the region highlighted by the dotted square. \textbf{(b)} Example image from the 3D stack taken \SI{1.5}{\micro\meter} above the bottom of the cell. \textbf{(c)} Corresponding YZ projection and \textbf{(d)} XZ projection shown at the coordinates associated with the nanodiamond indicated by the dotted lines in (b). Mitochondria are visible above the nanodiamond confirming internalization. Scalebars in (a)-(d) is \SI{3}{\micro\meter}. 
    \textbf{(e)} XY confocal image of a HeLa cell taken on the ODMR confocal microscope prior to a temperature measurement. \textbf{(f)} XZ confocal image of the same cell. Scalebars in (e) and (f) are \SI{5}{\micro\meter}. The blue curves represent the trajectory of the tracked nanodiamond and the white dashed curves represent the cell contour.
    XY is the imaging plane and Z is the optical axis direction in all Figures.
    }
    \label{fig:UptakeFigure}
\end{figure*}
We verify the nanodiamond internalisation by taking 3D confocal images of both the nanodiamonds and dyed mitochondria (Fig. \ref{fig:UptakeFigure} \textbf{(a)}). Nanodiamonds uptaken by the cell will be surrounded by the mitochondrial network in 3D (XY, YZ and XZ slices in Fig. \ref{fig:UptakeFigure} \textbf{(b)-(d)}). Before temperature measurements on the home-built confocal, we use the fluorescence from the cell culture medium to confirm the nanodiamond is in the cytoplasm in 3D (XY and XZ slices in Fig. \ref{fig:UptakeFigure} \textbf{(e), (f)}).

\section{Track segmentation based on the directionality ratio}
To identify segments of the nanodiamond trajectory that have a statistically significant directionality ratio, $\gamma$, we first consider the motion of a particle undergoing Brownian motion in two dimensions. After a total of $N$ steps, the distance travelled by the particle, $l'$, will be equal to
\begin{equation}
    l' = \sum_{t=1}^{N}\sqrt{\sum_{i=1}^{M}(\Delta r_{i,t})^2}
\end{equation}
where $\Delta r_{i,t} \sim \mathcal{N}(0,1)$ and $M$ = the number of dimensions (for a 2D projection, $M = 2$). The term inside the outer summation can be described by the variable, $Q_t \sim \chi(M)$, where $\chi(M)$ is the chi-distribution (namely the positive square root of the sum of squares of a set of independent random variables each following a standard normal distribution),
\begin{equation}
    l' = \sum_{t=1}^{N}Q_t.
\end{equation}
Now, using the central limit theorem, $l'$ can be expressed in terms of the normal distribution,
\begin{equation}
    \frac{l'}{\sqrt{N\sigma^2}} \sim \mathcal{N}(0,1) + \frac{\sqrt{N}\mu}{\sigma} 
\end{equation}
where $\mu$ and $\sigma$ are the mean and standard deviation for $\chi(M)$ respectively. For the same particle, the displacement, $d'$ can be described by,
\begin{equation}
    d' = \sqrt{\sum_{i=1}^{M}\left(\sum_{t=1}^{N}\Delta r_{i,t}\right)^2}
\end{equation}\\
which again can be characterised by the chi-distribution by normalizing by $\sqrt{N}$,
\begin{equation}
    \frac{d'}{\sqrt{N}} \sim \chi(M)
\end{equation}
The directionality ratio is defined as,
\begin{equation}
    \gamma = \frac{d'}{l'}.
\end{equation}\\
To find the probability density function for the directionality ratio, $f_\gamma$ we must define a new variable, $\eta$, such that,
\begin{equation}
    \eta = \frac{1}{\gamma}\sqrt{\frac{M}{\sigma^2}} \sim \frac{\mathcal{N}(0,1)+\frac{\sqrt{N}\mu}{\sigma}}{\frac{\chi(M)}{\sqrt{M}}}.
\end{equation}\\
As this form is indicative of a non-central t-distribution with $M$ degrees of freedom and a non-centrality parameter of $\frac{\sqrt{N}\mu}{\sigma}$, the probability density function for $\eta$, $f_\eta$ can be ascertained analytically. As the total integrals for $f_\gamma$ and $f_\eta$ must equal 1, the function of $f_\gamma$ can be found numerically,
\begin{equation}
    f_\gamma = \bigg\vert \frac{d\eta}{d\gamma} \bigg\vert f_\eta = \frac{1}{\gamma^2}\sqrt{\frac{M}{\sigma^2}}f_\eta.
\end{equation}\\
Regions of the nanodiamond trajectories were deemed statistically significant if the directionality ratio exceeded the critical value corresponding to a $\lt 5\%$ probability of occurring randomly for a particle undergoing Brownian motion. For the experimental data, $N$ = 75 (corresponding to 720 ms of tracking time), resulting in a critical value for $\gamma$ of 0.228, as shown in Fig. \ref{fig:DR_fig}.

\begin{figure*}[ht]
    \centering
    \includegraphics[width=\textwidth*2/5]
    {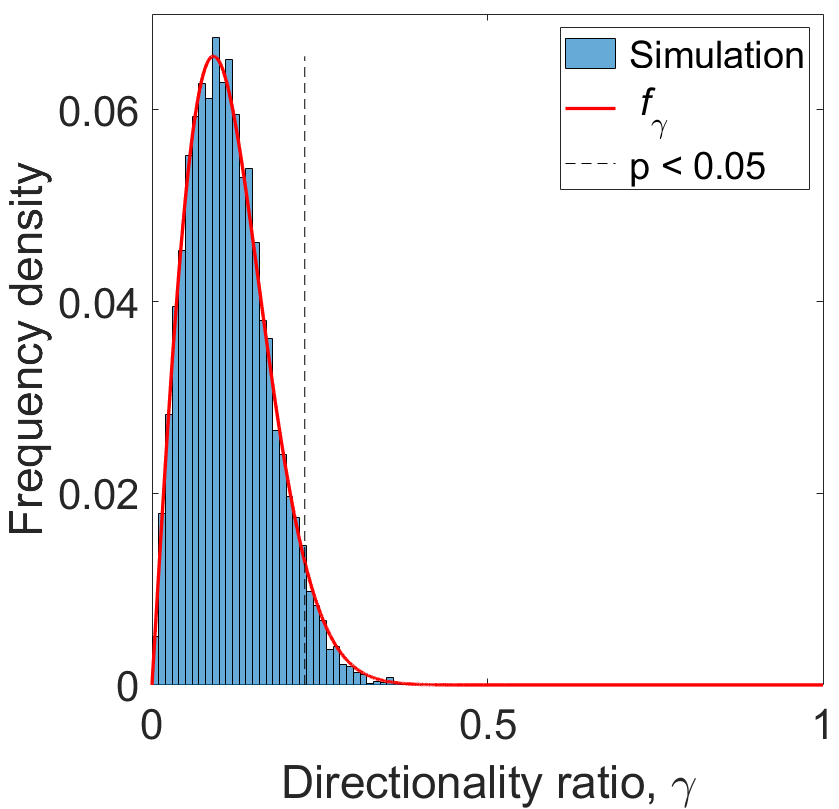}
    \caption{Extraction of critical dimensionality ratio value. For $N$ = 75, the directionality ratio of a particle undergoing Brownian motion in two dimensions at each time step over 10000 increments was simulated (blue bars) and matched with the probability density function, $f_\gamma$ as derived in the text. The critical value of 0.228 corresponds to the 95\% confidence interval for statistical significance.
    }
    \label{fig:DR_fig}
\end{figure*}
\noindent When identifying the trajectories of the nanodiamonds that were considered to be directed motion in cells, a length threshold of 500 nm was also used to remove displacements due to movement around or induced by organelles (for example the remodelling of the mitochondrial network).

\section{Deviation in extracellular and intracellular temperature}
In the main text we state that the intracellular temperature at the nanodiamond location over the timescale of the measurements equals the external temperature. Here, we give a detailed analysis of the statistical significance of this result.//
In response to an external temperature rise, $\Delta T_\mathrm{ext}$, the potential effects of physical thermal shielding and active cellular response could cause a different internal temperature rise, $\Delta T_\mathrm{int}$. This would cause the ODMR frequency of an internalised nanodiamond to shift by less (more) than $\kappa_\mathrm{NV}\Delta T_\mathrm{out}$. A smaller (larger) effective $\kappa_\mathrm{eff} = \kappa_\mathrm{NV}\Delta T_\mathrm{int}/\Delta T_\mathrm{ext} \approx \kappa_\mathrm{NV} +\Delta \kappa_\mathrm{eff}$ would be measured.  
\begin{figure*}[ht]
    \centering
    \includegraphics[width=\textwidth]{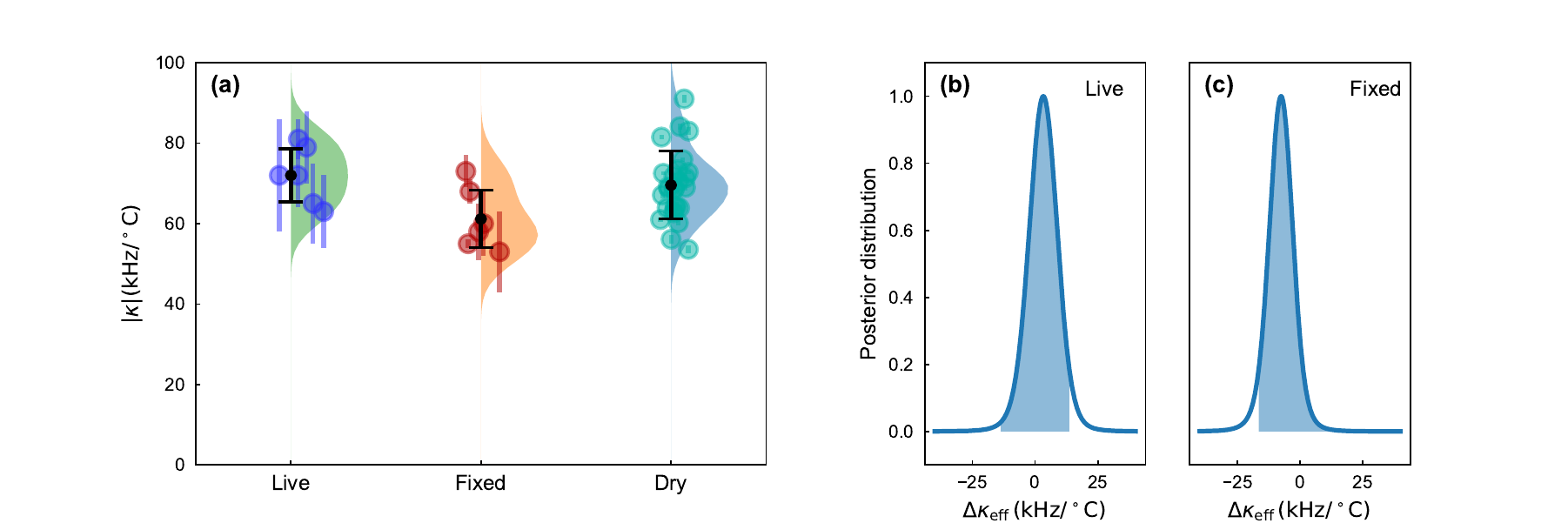}
    \caption{Distribution of $\kappa$ across nanodiamonds measured in different environments. \textbf{(a)} Variation of $\kappa$ across different experimental conditions: live cells (blue data), PFA fixed cells (red data) and dropcast on a dry substrate (cyan data). The shaded region corresponds to the inferred statistical distribution using the kernel density estimation method. The black errorbars indicate the mean and standard deviation of the corresponding data. \textbf{(b, c)} Posterior distribution of $\Delta\kappa_\mathrm{eff}$ for live and fixed cells given the prior distribution of dry sample variation and measurement uncertainty.
    }
    \label{fig:KappaViolinAndPosterior}
\end{figure*}\\
This measurement is in general confounded by the variation in $\kappa$ between nanodiamonds. For example, although the nanodiamond presented in the main text Fig. 2 \textbf{(b)} has a proportionality constant of $\kappa = -60.0(4)\,\mathrm{kHz/^{\circ} C}$, we measured a range from $-53.6(1)\,\mathrm{kHz/^{\circ} C}$ to $-91.1(1)\,\mathrm{kHz/^{\circ} C}$ in our samples of 26 nanodiamonds on dry substrates (cyan in Fig. \ref{fig:KappaViolinAndPosterior} \textbf{(a)}). If a constant value of $-74\,\mathrm{kHz/^{\circ} C}$ is assumed, this variation can incur a $\pm 25\%$ systematic error, which highlights the need for a per-diamond calibration. The large intrinsic variation also implies that the cell-to-cell variation may be overestimated if a constant value of $\kappa$ is assumed.\\
The intracellular response can be verified using statistical analysis. We measure the intracellular $\kappa$ for a series of six experiments (blue in Fig. \ref{fig:KappaViolinAndPosterior} \textbf{(a)}) in live cells and a series of six experiments in fixed cells (red in Fig. \ref{fig:KappaViolinAndPosterior} \textbf{(a)}). As there are intrinsic variations between nanodiamonds, the naive combination of uncertainties in measured values of intracellular $\kappa_\mathrm{eff}$ and $\kappa_{NV}$ is not sufficient to estimate that of $\delta \kappa_\mathrm{eff}$. As such we adopt a Bayesian approach similar to Ref. \cite{sotoma2021situ} to obtain the statistical distribution of $\delta \kappa_\mathrm{eff} = \kappa_\mathrm{eff} - \kappa_\mathrm{NV}$, which incorporates uncertainties due to cellular noise, measurement uncertainty and intrinsic diamond variations. On the timescale accessible to our experiments, $\delta \kappa_\mathrm{live} = 3(13)\,\mathrm{kHz/^{\circ} C}$ (Fig. \ref{fig:KappaViolinAndPosterior} \textbf{(b)}) and $\Delta \kappa_\mathrm{fixed} = -7(11)\,\mathrm{kHz/^{\circ}C}$ (Fig. \ref{fig:KappaViolinAndPosterior} \textbf{(c)}). These deviations are insignificant and therefore corroborate the results where the same cell is measured before and after fixation.

\section{Complex modulus analysis for nocodazole-treated cells}

\begin{figure*}[ht]
    \centering
    \includegraphics[width=\textwidth*2/5]{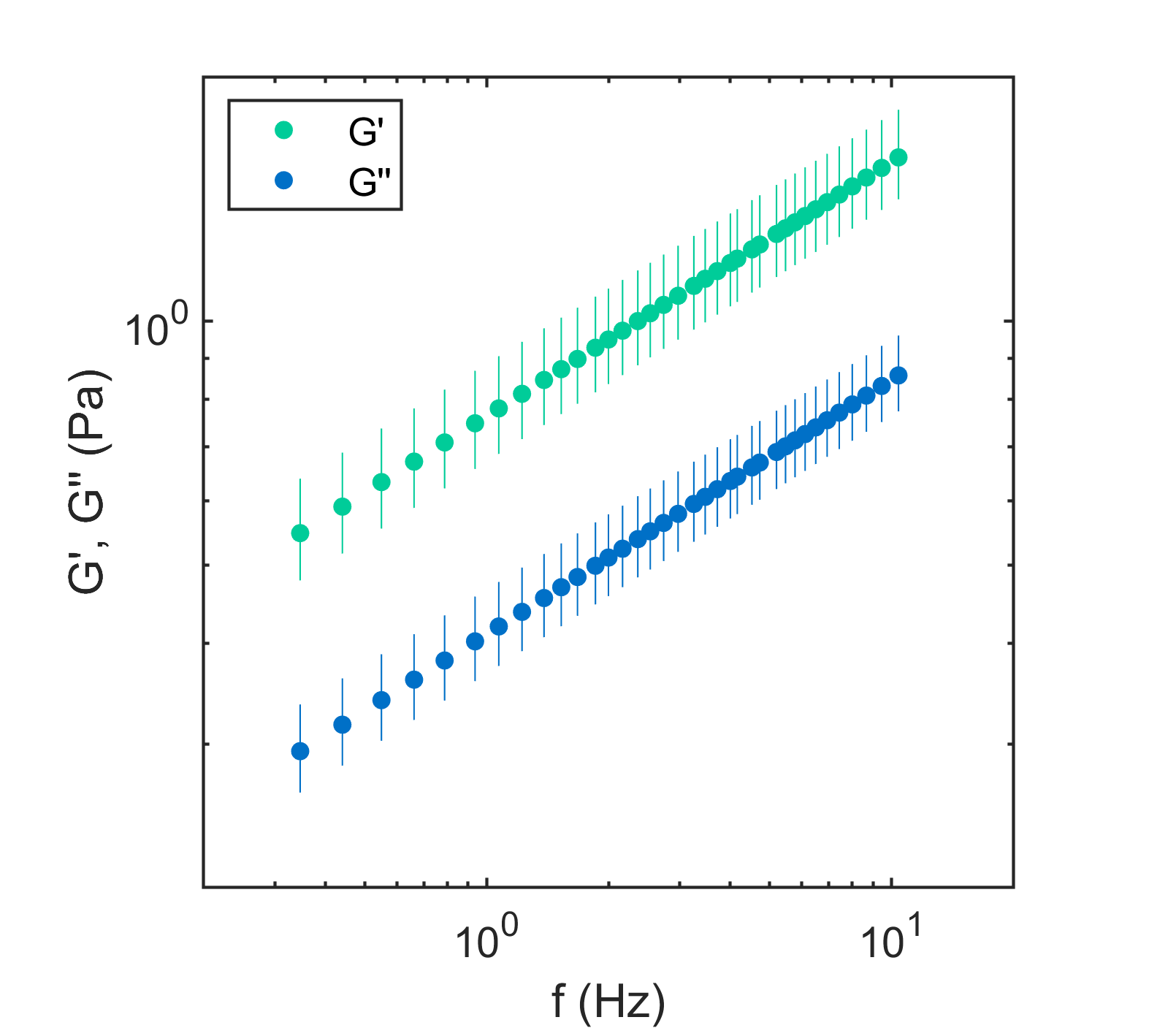}
    \caption{Complex modulus for nanodiamonds in nocodazole-treated cells, showing that the cytoplasm displays elasticity-dominated behaviour.} 
    \label{fig:gStarNoc}
\end{figure*}

Using the trajectories of nanodiamonds in HeLa cells subjected to a 1 hour treatment of 50 $\mu$M nocodazole, the intracellular material properties were investigated using the complex modulus, $G^*$. In Fig. \ref{fig:gStarNoc}, the average elastic and viscous components of $G^*$ ($G'$ and $G''$ respectively) are shown, with the elastic component dominating.


\end{document}